\documentclass[aps,pra,10pt,twocolumn]{revtex4}

\input{Qcircuit}
\usepackage{graphicx,graphics,wrapfig,rotating}         
\usepackage{bm,fancybox}
\usepackage{amsmath,amssymb}                          
\usepackage{times,euscript,oldgerm}              %
\usepackage[english]{babel}                      %
\usepackage{psfrag}
\usepackage{color}
\usepackage{epsfig}
\usepackage{dcolumn}

\newcommand{\up}{\ensuremath{\left|\uparrow\right\rangle}}
\newcommand{\down}{\ensuremath{\left|\downarrow\right\rangle}}

\begin{document}

\title{Large Scale Modular Quantum Computer Architecture  \\ with Atomic Memory and Photonic Interconnects}
\author{C. Monroe$^1$, R. Raussendorf$^2$, A. Ruthven$^2$, K. R. Brown$^3$, P. Maunz$^{4*}$, L.-M. Duan$^5$, and J. Kim$^4$}

\affiliation{$^1$ Joint Quantum Institute, University of Maryland Department of Physics and  
                  National Institute of Standards and Technology, College Park, MD  20742, USA \\
             $^2$ Department of Physics and Astronomy, University of British Columbia, Vancouver, BC V6T1Z1, Canada \\
             $^3$ Schools of Chemistry and Biochemistry; Computational Science and Engineering; and Physics, 
                  Georgia Institute of Technology, Atlanta, GA 30332, USA \\
             $^4$Department of Electrical and 
                  Computer Engineering, Duke University, Durham, NC  27708, USA \\
             $^5$ Department of Physics and MCTP, University of Michigan, Ann Arbor, MI 48109, USA and 
                  Center for Quantum Information, Tsinghua University, Beijing 100084, China}
\altaffiliation[Present address: ]{Sandia National Laboratories, Albuquerque, N.M. 87123, USA}
\date{\today}

\begin{abstract}
The practical construction of scalable quantum computer hardware capable of executing non-trivial quantum algorithms will require the juxtaposition of different types of quantum systems. We analyze a modular ion trap quantum computer architecture with a hierarchy of interactions that can scale to very large numbers of qubits. Local entangling quantum gates between qubit memories within a single register are accomplished using natural interactions between the qubits, and entanglement between separate registers is completed via a probabilistic photonic interface between qubits in different registers, even over large distances.  We show that this architecture can be made fault-tolerant, and demonstrate its viability for fault-tolerant execution of modest size quantum circuits.
\end{abstract}

\maketitle 

\section{Introduction}
A quantum computer is composed of at least two quantum systems that serve critical functions: a reliable quantum memory for hosting and manipulating coherent quantum superpositions, and a quantum bus for the conveyance of quantum information between memories.  Quantum memories are typically formed out of matter such as individual atoms, spins localized at quantum dots or impurities in solids, or superconducting junctions \cite{QC}.  On the other hand, the quantum bus typically involves propagating quantum degrees of freedom such as electromagnetic fields (photons) or lattice vibrations (phonons).  A suitable and controllable interaction between the memory and the bus is necessary to efficiently execute a prescribed quantum algorithm.  The current challenge in any quantum computer architecture is to scale the system to very large sizes, where errors are typically caused by speed limitations and decoherence of the quantum bus or its interaction with the memory. The most advanced quantum bit (qubit) networks have thus been established only in very small systems, such as individual atomic ions bussed by the local Coulomb interaction \cite{WinelandBlatt08} or superconducting Josephson junctions coupled capacitively or through microwave striplines \cite{sup1, sup2}.  In this paper, we propose and analyze a hierarchy of quantum information processing units in a modular quantum computer architecture that may allow the scaling of high performance quantum memories to useful sizes~\cite{MonroeScience2013}.  This architecture compares to the ``multicore" classical information processor, and is suitable for the implementation of complex quantum circuits utilizing the flexible connectivity provided by a reconfigurable photonic interconnect network. Unlike previous related proposals 
\cite{Thaker2006, Jiang07, MoehringJOSA07, Helmer2009, NV-arch, Fujii2012}, we show this reconfigurable architecture can be made fault-tolerant over a wide range of system parameters, using a variety of fault-tolerant schemes.  All of the rudiments of this architecture have been demonstrated in small-scale trapped ion systems, and we speculate on the technological hurdles ahead in order to realize such a system. 

We specialize to the use of atomic ion qubit memories, due to the outstanding qubit properties demonstrated to date. Qubits stored in ions enjoy a level of coherence that is unmatched in any other physical system, underlying the reason such states are also used as high performance atomic clocks.  Moreover, atomic ions can be initialized and detected with nearly perfect accuracy using conventional optical pumping and state-dependent fluorescence techniques~\cite{NIST}. There have been many successful demonstrations of controlled entanglement of several-ion quantum registers in the past decade involving the use of qubit state-dependent forces supplied by laser beams \cite{WinelandBlatt08,Haffner08}.  These experiments exploit the collective motion of a small number of trapped ion qubits, but as the size of the ion chain grows, such operations are more susceptible to external noise, decoherence, or speed limitations.

One promising approach to scaling trapped ion qubits is the quantum charge-coupled device (QCCD), where physical shuttling of ions between trapping zones in a multiplexed trap is used to transfer qubits between (short) chains of ions~\cite{QCCD, NIST}. This approach involves advanced ion trap structures, perhaps with many times more discrete electrodes as trapped ion qubits, and therefore motivates the use of micrometer-scale surface traps \cite{NISTsurface05, seidelin06, WangSurfGate10} and novel fabrication techniques \cite{Kim05, Sandia, GTRI}. The shuttling approach requires careful control of the time-varying trapping potential to manipulate the position of the atomic ion, and cannot easily be extended over large distances for quantum communications applications. The QCCD approach is expected to enable a quantum information processing platform where basic quantum error correction and quantum algorithms can be realized. Further scaling is likely limited by the complexity of the trap design, diffraction of optical beams, and the hardware controllers to operate the system.

\begin{figure}
\includegraphics[width=1.0\linewidth]{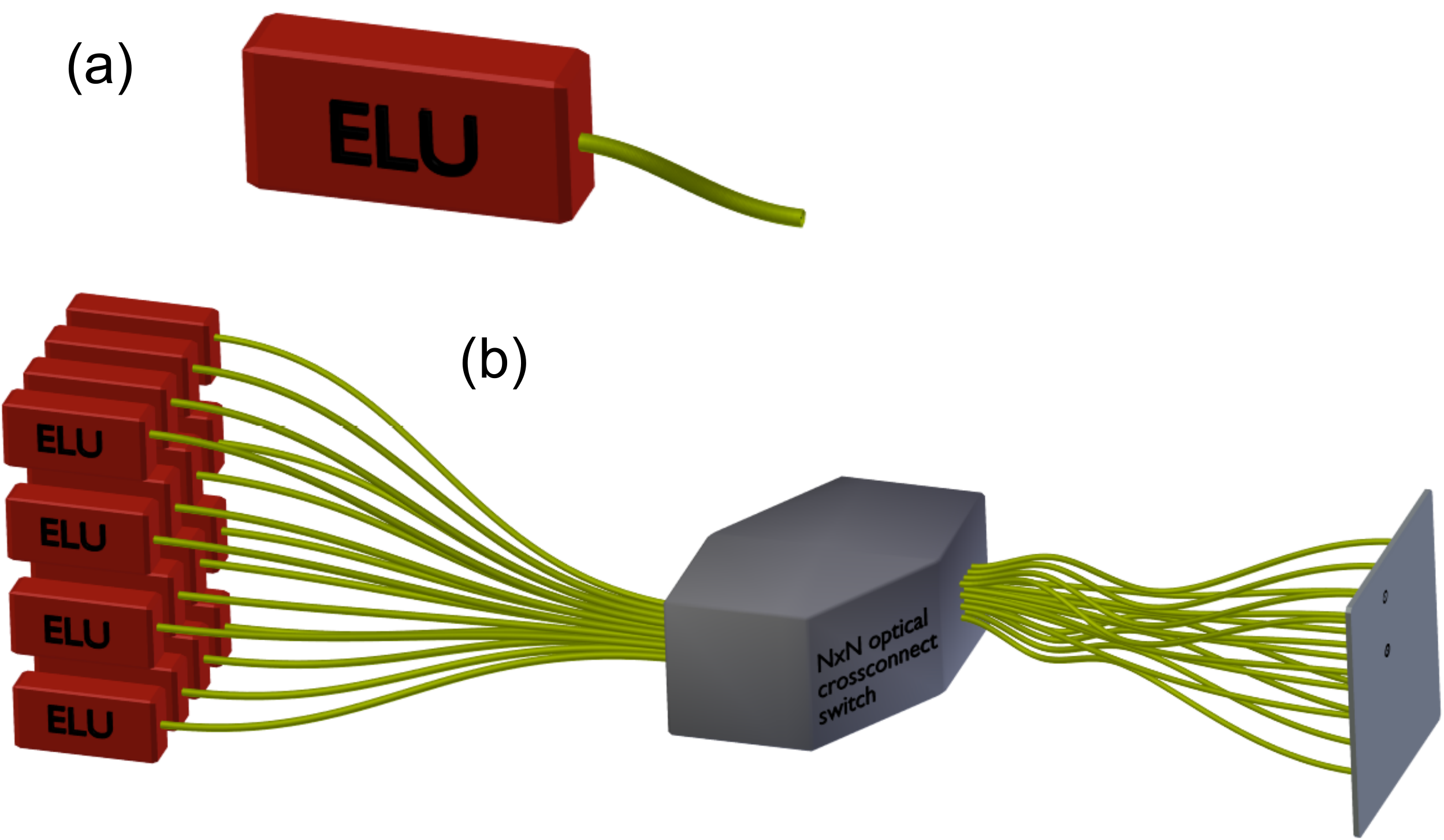}
\caption{(Color online) Hierarchical modular quantum computer architecture hosting $N=N_{ELU}N_q$ qubits. 
(a) The elementary logic units (ELU)
consists of a register of $N_q$ trapped atomic ion qubits, whereby entangling
quantum logic gates are mediated through the local Coulomb interaction between qubits.  
(b) One or more atomic qubits within each of the $N_{ELU}$ registers are coupled to photonic quantum channels, and through a 
reconfigurable optical crossconnect switch (OXC, center), fiber beamsplitters and position 
sensitive imager (right), qubits between different registers can be entangled.}
\label{MUSIQC}
\end{figure}

Here we describe and analyze a modular universal scalable ion trap quantum computer (MUSIQC) architecture that may enable construction of quantum processors with up to $10^6$ qubits utilizing component technologies that have already been demonstrated.  
This architecture features two elements: stable trapped ion multi-qubit registers that can further be connected with ion shuttling, and scalable photonic interconnects that can link these registers in a flexible configuration over large distances, as shown in Fig. \ref{MUSIQC}.  We articulate architectural advantages of this approach that allows significant speedup and resource reduction in quantum circuit execution over other hardware architectures, enabled by the ability to operate quantum gates between qubits throughout the entire processor regardless of their relative location. Finally, we prove how such a quantum network can support fault-tolerant error correction even in the face of probabilistic interconnects, and discuss the technological developments necessary for its realization. While we focus our discussions on quantum registers composed of trapped atomic ions, the networking aspect of this architecture is applicable to other qubit platforms that feature strong optical transitions, such as quantum dots, neutral atoms, or nitrogen-vacancy (NV) color centers in diamond~\cite{QC}.

\section{Quantum Computing in a Modular Architecture}

\subsection{The Modular Elementary Logic Unit (ELU)}
\begin{figure}
\includegraphics[width=1.0\linewidth]{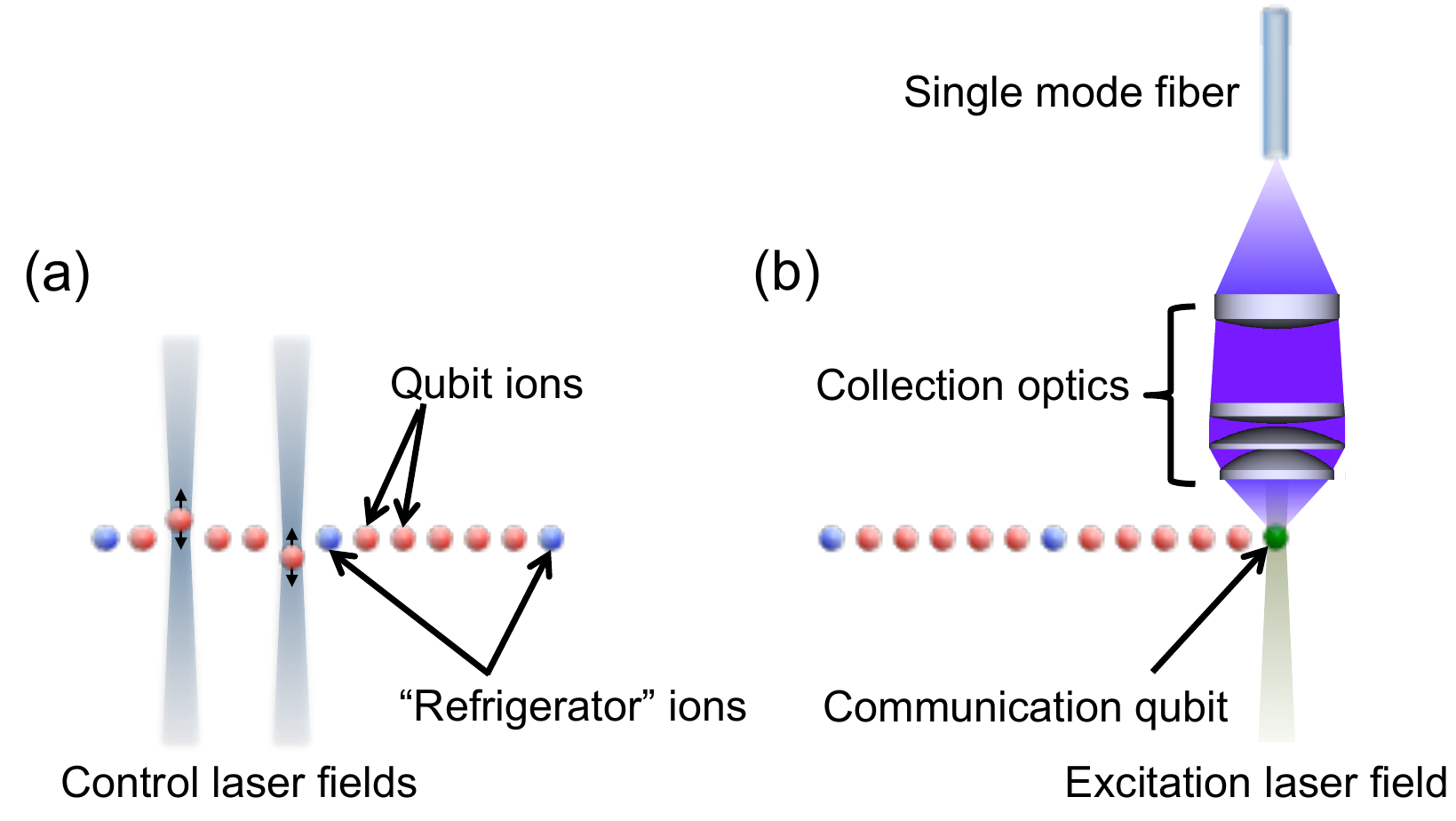}
\caption{(Color online) Elementary Logic Unit (ELU) composed of a single crystal of $N_q$ trapped atomic ion 
qubits coupled through their collective motion.  (a) Classical laser fields impart qubit
state-dependent forces on one or more ions, affecting entangling quantum gates
between the memory qubits. Second ion species is introduced as refrigeration ions. (b) One or more of the ions (rightmost in the figure) are coupled 
to a photonic interface, where a classical laser pulse maps the state of these communication qubits onto the states of single photons (e.g., polarization or frequency), which then propagate along an optical fiber to be interfaced with other ELUs.}
\label{ELU}
\end{figure}

The base unit of MUSIQC is a collection of $N_q$ qubit memories with local interactions, called 
the Elementary Logic Unit (ELU).  Quantum logic operations within the ELU are ideally fast and deterministic, with error rates sufficiently small that fault-tolerant error correction within an ELU is possible \cite{MikeAndIke}. 
We represent the ELU with a crystal of $N_q \gg 1$ trapped atomic ions as shown in Fig. \ref{ELU}a, with each 
qubit comprised of internal energy levels of each ion, labeled as \up and \down, separated by 
frequency $\omega_0$.  We assume the qubit levels are coupled through an atomic dipole operator $\hat{\mu} = \mu(\up\bra{\downarrow} + \down\bra{\uparrow})$.  The ions interact through their external collective modes of quantum harmonic motion.  Such phonons can be used to mediate entangling gates through application of qubit-state-dependent optical or microwave dipole forces \cite{CZ,MS, rfgate}. There are many known protocols for phonon-based gates between ions, and here we summarize the main points relevant to the size of the ELU and the larger architecture.

An externally applied near-resonant running wave field with amplitude $E(\hat{x}) = E_0 e^{ik\hat{x}}$ and 
wavenumber $k$ couples to the atomic dipole through the interaction Hamiltonian $\hat{H} = -\hat{\mu}E(\hat{x})$, and 
by suitably tuning the field near sidebands induced by the harmonic motion of the ions \cite{NIST} 
a qubit state dependent force results.  In this way, qubits can be 
mapped onto phonon states \cite{CZ, NIST} and then onto other qubits for entangling operations with characteristic speed $R_{\text{gate}} = \eta\Omega$, where $\eta = \sqrt{\hbar k^2/(2m_0 N_q\omega)}$ is the Lamb Dicke parameter, $m_0$ is the mass of each ion, $\omega$ the frequency of harmonic oscillation of the collective phonon mode, 
and $\Omega = \mu E_0 /2\hbar$ is the Rabi frequency of the atomic dipole independent of motion. 
For optical Raman transitions between qubit states (e.g., atomic hyperfine ground states) \cite{NIST}, 
two fields are each detuned by $\Delta$ from an excited state of linewidth $\gamma \ll \Delta$, and 
when their difference frequency is near resonant with the qubit frequency splitting $\omega_0$, 
we use instead $\Omega = (\mu E_0)^2/(2\hbar^2\Delta)$.

The typical gate speed within an ELU therefore slows down with the number of qubits $N_q$ as $R_{\text{gate}} \sim N_q^{-1/2}$.  As the size of the ELU grows, so will the coupling between the modes of collective motion that could lead to crosstalk.  However, through the use of pulse-shaping techniques \cite{PulseShaping}, the crosstalk errors need not be debilitating, although the effective speed of a gate will slow down with size $N_q$. Changes of the ions' motional states during the gate, arising from sources like heating of the motional modes~\cite{Turchette,DeslauriersHeat,Lab08b} or fluctuating fields, will degrade the quality of the gates, leading to practical limits on the size of the ELU on which high performance gates can be realized. It is likely that long chains will require periodic ``refrigerator'' ions to remove motional excitations between gates. Since cooling is a dissipative process, these cooling ions should be chosen to be different isotope or species of ions and quench motional heating through sympathetic cooling~\cite{Lin2011}. We estimate that ELUs ranging from $N_q=10-100$ should be possible \cite{WinelandBlatt08,Haffner08}. More than one ELU chain can be integrated into a single chip by employing ion shuttling through more complex ion trap structures \cite{QCCD}. Such extended ELUs (EELUs) consisting of $N_{E}$ ELU chains can contain a total of $N_q N_E = 20-1,000$ physical qubits. For simplicity, we focus the remainder of the article on systems with one ELU per chip ($N_E=1$).

\subsection{Probabilistic Linking of ELUs}
Two qubits from a pair of ELUs (or EELUs) can be entangled by each emitting photons that interfere with each other. Entanglement generated between these ``communication qubits'' can be utilized as a resource to perform a two-qubit gate between any pair of qubits, one from each ELU, using local qubit gates, measurements, and classical communication between the ELUs. In this scheme, the communication qubit is driven to an excited state with fast laser pulses whose duration $\tau_e \ll 1/\gamma$, so that no more than
one photon is emitted from each qubit per excitation cycle following the atomic radiative selection rules (Fig. \ref{ELU}b). The photon can be post-selected so that one of its degrees of freedom (polarization, frequency, etc.) is entangled with the state of the communication qubit \cite{BlinovNature2004,ToganNature2010,DeGreveNature2012,GaoNature2012}. When the photons from two communication qubits are mode-matched and interfere on a $50/50$ 
beamsplitter, detectors on the output modes of the beamsplitter can 
herald the creation of entanglement between the memory qubits \cite{Cabrillo99,Duan03,Simon2003,Moehring07,DuanRMP}.

\begin{figure}
\includegraphics[width=1.0\linewidth]{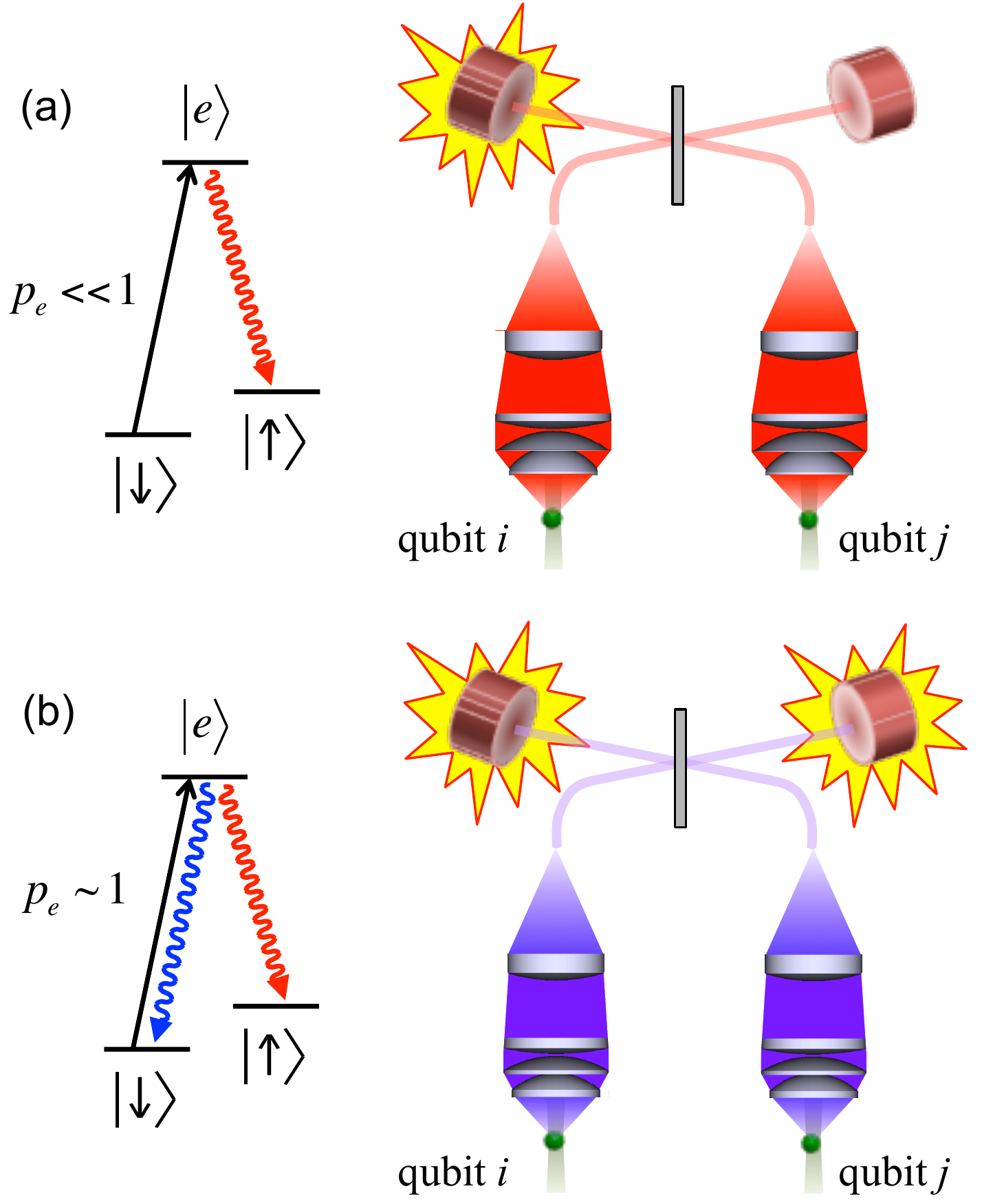}
\caption{(Color online) (a) Type I interference from photons emitted from two communication qubits. Each qubit is weakly excited so that single photon emission has a very small probability yet is correlated with the final qubit state.  The output photonic channels are mode-matched with a 50/50 beamsplitter and subsequent detection of a photon from either output port heralds the entanglement of the communication qubits.  The probability of two photons present in the system is much smaller than that of detecting a single photon.  (b) Type II interference involves the emission of one photon from each communication qubit, where the internal state of the photon (e.g. its color) is correlated with the qubit state.  After two photon interference at the beamsplitter, coincidence detection of photons at the two detectors heralds the entanglement of the communication qubits.}
\label{Interference}
\end{figure}

We consider two types of photonic connections, characterized by the number of total photons used in the entanglement 
protocol between two communication qubits \cite{DuanQIC}.  
For type I connections (shown in Fig. \ref{Interference}a), each communication qubit with an index $i$ (or $j$) is weakly excited with 
probability $p_e \ll 1$ and the state of the ion+photon qubit pair is approximately written (ignoring the higher-order excitation probabilities) as $\sim \sqrt{1-p_e}\down_i\ket{0}_i + e^{ikx_i}\sqrt{p_e}\up_i\ket{1}_i$ where $\ket{n}_i$ denotes the 
state of $n$ photons radiating from the communication qubit into an optical mode $i$, $x_i$ is the path length from the 
emitter $i$ to a beamsplitter, and $k$ the optical wavenumber \cite{Cabrillo99}.  
When two communication qubits $i$ and $j$ are excited in this way and the photons interfere at the beam splitter, 
the detection of a single photon in either detector placed at the two output ports of the beamsplitter heralds the creation of the state
$[e^{ikx_j}\down_i\up_j \pm e^{ikx_i}\up_i\down_j]/\sqrt{2}$ with success probability $p = p_e F\eta_D$, 
where $F$ is the fractional solid angle of emission collected, $\eta_D$ is the detection
efficiency including any losses between the emitter and the detector, and the sign in this state is determined by which one of the two detectors fires.  
Following the heralding of a single photon, the (small) probability of errors from double excitation and detector dark counts are given respectively by $p_e^2$ and $R_{\text{dark}}/\gamma$ where $R_{\text{dark}}$ is the rate of detector dark counts.  For type I connections to be useful, the relative optical path length $x_i - x_j$ must be stable to much better than the optical wavelength $\sim 2\pi/k$.  

For type II connections (shown in Fig. \ref{Interference}b), each communication qubit is excited with near unit probability $p_e \sim 1$ 
and the single photon carries its qubit through two distinguishable internal photonic states 
(e.g., polarization or optical frequency).  For example, the state of the system containing both communication and photonic qubits is written as
$[e^{ik_{\downarrow}x_i}\down_i\ket{\nu_{\downarrow}}_i + e^{ik_{\uparrow}x_i}\up_i\ket{\nu_{\uparrow}}_i]/\sqrt{2}$, where
$\ket{\nu_{\downarrow}}_i$ and $\ket{\nu_{\uparrow}}_i$ denote the frequency qubit states of a single-photon emitted by the $i$-th communication qubit with wavenumbers $k_{\downarrow}$ and $k_{\uparrow}$ associated with optical frequencies $\nu_{\uparrow}$ and $\nu_{\downarrow}$, respectively. Here, $|\nu_{\uparrow}-\nu_{\downarrow}| = \omega_0 \gg \gamma$ so that these two frequency qubit states are distinguishable. The coincidence detection of photons from two such communication qubits $i$ and $j$ after interfering at a 50/50 beam splitter herald the successful entanglement of the communication qubits, creating the state $[e^{i(k_{\downarrow}x_i+k_{\uparrow}x_j)}\down_i\up_j -e^{i(k_{\uparrow}x_i+k_{\downarrow}x_j)}\up_j\down_i]/\sqrt{2}$ with success probability $p = (p_e F\eta_D)^2/2$ \cite{Duan03,Simon2003}. 

The success probability of the 2-photon type II connection may be lower than that of the type I connection when the light collection efficiency is low, but type II connections are much less sensitive to optical path length fluctuations. The stability requirement of the relative path length $x_i-x_j$ is only at the level of the wavelength associated with the difference frequency $2\pi c/\omega_0$ of the photonic frequency qubit, which is typically at the centimeter scale for hyperfine-encoded communication qubits. 

In both cases, the mean connection time is given by $\tau_E = 1/(Rp)$ where $R$ is the repetition rate of the 
initialization/excitation process and $p$ is the success probability of generating the entanglement.  For atomic transitions, $R \sim 0.1 (\gamma/2\pi)$, and for typical free-space light collection $(F \sim 10^{-2}$) and taking $\eta_D \sim 0.2$, we find for a type I connection 
$\tau_E \sim 5$ msec ($p_e = 0.05$) and for a type II connection $\tau_E \sim 250$ msec  where we have assumed $\gamma/2\pi = 20$ MHz. Type II connections eventually outperform that of type I with more efficient light collection, which can be accomplished by integrating optical elements with the ion trap structure without any fundamental loss in fidelity \cite{KimMaunzKimPRA2011}. Eventually, $\tau_E$ lower than 1 msec should be possible in both types of connections.

The process to generate ion-ion entanglement using photon interference requires resonant excitation of the communication qubits, and steps must be taken to isolate the communication qubit from other memory qubits so that scattered light from the excitation laser and the emitted photons do not disturb the spectator memory qubits.  It may be necessary to physically separate or shuttle the communication qubit away from the others, invoking some of the techniques from the QCCD approach. This crosstalk can also be eliminated by utilizing a different atomic species for the communication qubit \cite{Schmidt05}, so that the excitation and emitted light is sufficiently far from the memory qubit optical resonance to avoid causing decoherence.  The communication qubits do not require excellent quantum memory characteristics, because once the entanglement is established between the communication qubits in different ELUs, they can immediately be swapped with neighboring memory qubits in each ELU.

\subsection{Reconfigurable Connection Network in MUSIQC}
The MUSIQC architecture allows a large number $N_{ELU}$ of ELUs (or EELUs) to be connected with each other using the photonic channels, as shown in Fig. \ref{MUSIQC}. The connection is made through an optical crossconnect (OXC) switch with $N_{ELU}$ input and output ports. The photon emitted from the communication qubit in each ELU is collected into a single-mode fiber and directed to a corresponding input port of the OXC switch. Up to $N_{ELU}/2$ Bell state detectors, each comprised of two fibers interfering on a beam splitter and two detectors, are connected to the output ports of the OXC switch. The OXC switch is capable of providing an optical path between any input fiber to any output fiber that is not already connected to another input fiber. An ideal OXC switch achieves full non-blocking connectivity with uniform optical path lengths. This optical network provides fully reconfigurable interconnect network for the photonic qubits, allowing entanglement generation between any pair of ELUs in the processor with up to $N_{ELU}/2$ such operations running in parallel. OXC switches that support $200-1,100$ ports utilizing micro-electromechanical systems (MEMS) technology have been constructed and are readily available~\cite{KimPTL2003,NeilsonJLT2004}.  In practice, the photon detection can be accomplished in parallel with a conventional charge-coupled-device (CCD) imager or an array of photon-counting detectors, with pairs of regions on the CCD or the array elements associated with particular pairs of output ports from the fiber beamsplitters, as shown in Fig. \ref{MUSIQC}.  

\section{Performance Advantage of MUSIQC Architecture}

\subsection{Computation Model in MUSIQC}
In the circuit model of quantum computation, execution of two-qubit gates creates the entanglement necessary to exploit the power of quantum physics in computation \cite{MikeAndIke}. In the alternate model of measurement-based cluster-state quantum computation, all of the entanglement is generated at the beginning of the computation, followed by conditional measurements of the qubits \cite{Raussendorf01}. The MUSIQC architecture presented here follows the circuit model of computation within each ELU, but the probabilistic connection between ELUs is carried out by generation of entangled Bell pairs similar to the cluster-state computation model. In this sense, MUSIQC realizes a hybrid model of quantum computation, driven by the generation rate and burn rate of entanglement between the ELUs. In the event the generation rate of entangled Bell pairs between ELUs is lower than the burn rate, each ELU would require the capacity to store enough initial entanglement so that the end of the computation can be reached at the given generation and burn rates of entanglement. The hybrid nature of MUSIQC provides a unique hardware platform with three distinct advantages: naturally parallel operation of each ELU, constant timescale to perform operations between distant qubits, and moderate ELU size adequate for practical implementation. One can further reduce the entanglement generation time by time-division multiplexing (TDM) the communication ports at the expense of added qubits. Moreover, the temporal mismatch between the remote entanglement generation and local gates is reduced as the requirement of error correction increases the logical gate time.

For complex quantum algorithm associated with a problem size of $n$ bits, logical operations between spatially distant qubit pairs are necessary. In a hardware architecture where only local gate operations are allowed ({\em e.g.}, nearest neighbor gates), performing gate operations between two (logical) qubits separated by long distances could lead to communication times polynomial in the distance between qubits, $O(n^k)$. When a large number of parallel operations is available, one can employ a nested entanglement swapping protocol to efficiently distribute entanglement with communication times scaling only logarithmically as a function of communication distance \cite{BriegelPRL1998}. The procedure requires extra qubits used to construct quantum buses for long-distance entanglement distribution, and architecture adopting such buses was referred to as the Quantum Logic Array (QLA) \cite{MetodiMicro2005}. We construct a simple model that provides a direct comparison between the QLA and MUSIQC architectures in terms of the resources required to execute useful quantum algorithms. Despite the slow entanglement generation times, we find that the performance of MUSIQC architecture is comparable to QLA (and its variations \cite{Metodi.T:2008ab}), with substantial advantage in required resources and feasibility for implementation. 

In our simplified model, we consider (1) hardware capable of implementing a Steane [[7,1,3]] quantum error correction code to multiple levels of concatenation, (2) where all gate operations are performed following fault-tolerant procedures. This simplified model is designed to estimate the execution time of the circuits in select architectures, and not intended to provide the complete fault-tolerant analysis of the quantum circuit. For this analysis, we therefore require that the physical error levels are sufficiently low ($\sim 10^{-7}$) to produce the correct answer with order-unity probability using only up to three levels of concatenation of Steane code. The hardware is based on trapped ion quantum computing with the assumptions for the timescales for quantum operation primitives summarized in Table \ref{Timescales}. The details of fault-tolerant implementation of universal gate set utilized in this analysis is summarized in Appendix~\ref{UniversalFTGates}.

\begin{table*}
\caption{Assumptions on the timescales of quantum operation primitives used in the model.}
\begin{ruledtabular}
\begin{tabular}{c | c c c c c} 
Quantum & Single-Qubit & Two-Qubit & Toffoli & Qubit & Remote Entanglement \\
Primitive & Gate & Gate & Gate & Measurement & Generation \\ \hline
Operation Time ($\mu$s) & 1  & 10  & 10  & 30  & 3000  \\ 
\end{tabular}
\end{ruledtabular}
\label{Timescales}
\end {table*}

\subsection{Construction of Efficient Arithmetic Circuits}

The example quantum circuit we analyze is an adder circuit that computes the sum of two $n$-bit numbers. Simple adder circuits form the basis of more complex arithmetic circuits, such as the modular exponentiation circuit that dominates the execution time of Shor's factoring algorithm \cite{VedralPRA1996}. Quantum adder circuits can be constructed using $X$, $CNOT$ and Toffoli gates. When only local interactions are available without dedicated buses for entanglement distribution, a quantum ripple-carry adder (QRCA) is the adequate adder of choice \cite{CuccaroArXiv2004}, for which the execution time goes as $O(n)$. For QLA and MUSIQC architectures, one can implement quantum carry-lookahead adder (QCLA) that is capable of reducing the runtime to $O$(log $n$) \cite{DraperQIC2006,VanMeterPRA2005}, at the expense of extra qubits and parallel operations. QCLA dramatically outperforms the QRCA for $n$ above $\sim 100$ in terms of execution time. Practical implementation of large-scale QCLAs are hindered by the requirement of executing Toffoli gates among qubits that are separated by long distances within the quantum computer. MUSIQC architecture flattens the communication cost between qubits in different ELUs, providing a suitable platform for implementing QCLAs. Alternatively, QLA architecture can also efficiently execute QCLAs using dedicated communication bus that reduces the connection time between two qubits (defined as the time it takes to generate entangled qubit pairs that can be used to teleport one of the qubits or the gate itself) to increase only as a logarithmic function of the separation between them~\cite{MetodiMicro2005}.

\subsection{MUSIQC Implementation}
In order to implement the QCLA circuit in MUSIQC architecture, each ELU should be large enough to accommodate the generation of the $\ket{\phi_+}_L$ state shown in Fig.~\ref{FTToffoli}a. This requires a minimum of 3 logical qubits and a 7-qubit cat state, and sufficient ancilla qubits to support the state preparation. We balance the qubit resource requirements with computation time by requiring four ancilla qubits per logical qubit, so that the 4-qubit cat states necessary for the stabilizer measurement can be created in parallel. Implementation of each Toffoli gate is realized by allocating a fresh ELU and preparing the $\ket{\phi_+}_L$ state, then teleporting the three qubits from other ELUs into this state. Once the gate is performed, the original logical qubits from the other ELUs are freed up and become available for another Toffoli gate. We find that $6n$ logical qubits placed on $6n/4 = 1.5n$ ELUs is sufficient to compute the sum of two $n$-bit integers using the QCLA circuit at the first concatenation level of Steane code encoding.

Teleportation of qubits into the ELU containing the prepared $\ket{\phi_+}_L$ state requires generation of entangled states via photon interference. In order to minimize the entanglement generation time, one should provide at least three optical ports to connect to these ELUs in parallel. In order to successfully teleport the gate, we need to create seven entangled pairs to each ELU holding the input qubits. The entanglement generation time can be reduced by running multiple optical ports to other ELUs in parallel (we call this the {\em port multiplexity} $m_p$). In a typical entanglement generation procedure, the ion is prepared in an initial state, and then excited using a short pulse laser ($\sim$5ps). The ion emits a photon over a spontaneous emission lifetime ($\sim$10ns), and the photon detection process will determine whether the entanglement generation from a pair of such ions is successful. If the entanglement generation is successful, the pair is ready for use in the computation. If not, the ions will be re-initialized ($\sim1\mu$s) and the process is repeated. Since the initialization time of the ion is $\sim$100 times longer than the time a photon is propagating in the optical port, one can utilize multiple ions per optical port and ``pipeline'' the photon emission process. In this time-division-multiplex (TDM) scheme, another ion is brought into the optical port to make another entanglement generation attempt while the initialization process is proceeding for the unsuccessful ion. This process can be repeated $m_T$ times using as many extra ions, before the first ion can be brought back (we call $m_T$ the {\em TDM multiplexity}). Using the port and TDM multiplexity, we can reduce the entanglement generation time by a factor of $m_p m_T$.

\begin{figure}
\includegraphics[width=8.5cm]{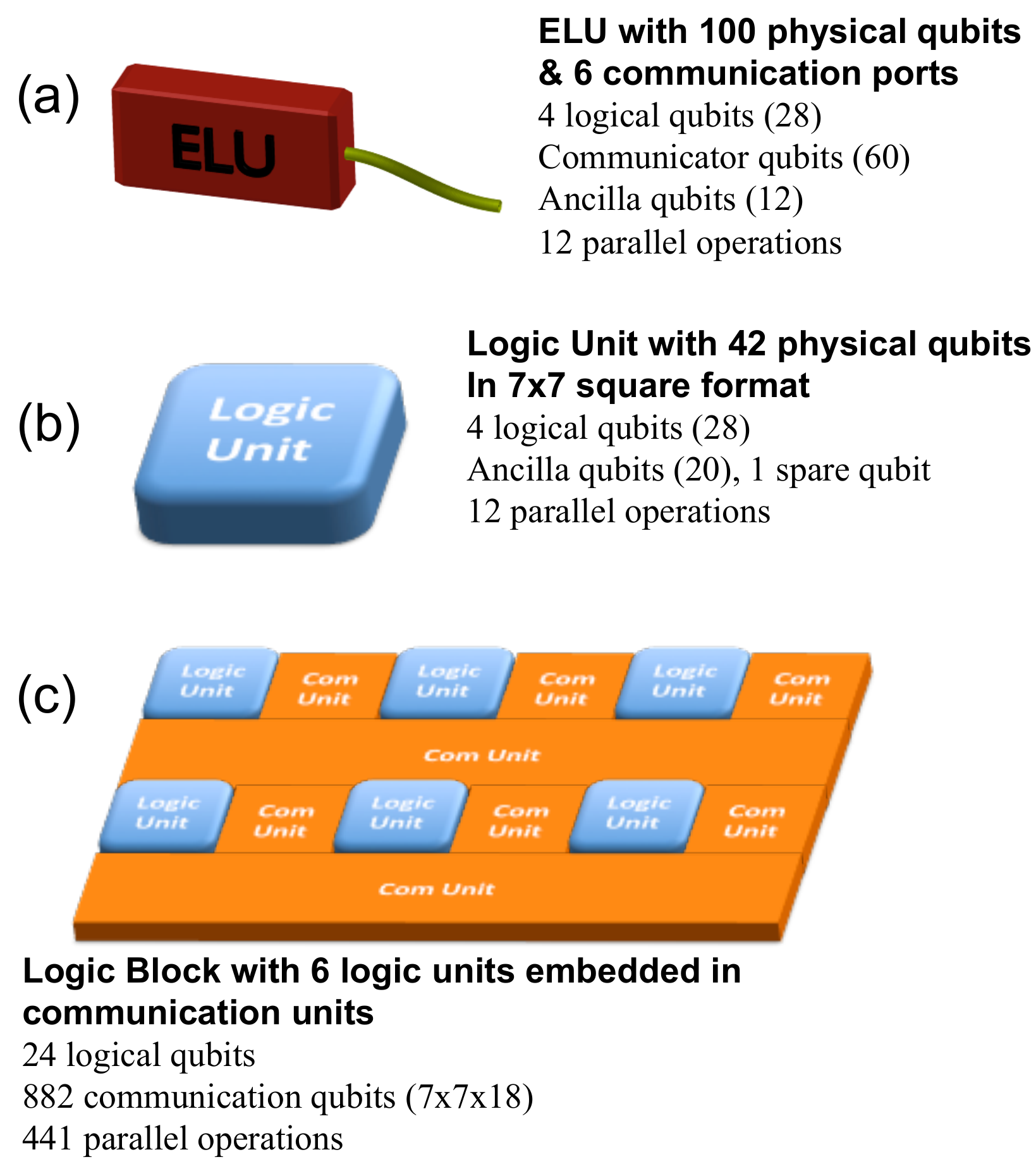}
\caption{(Color online) Example of the MUSIQC and QLA hardware considered. (a) Each ELU in MUSIQC is made up of 100 physical qubits and 6 communication ports (only one shown in the figure), where 60 qubits are used to increase the bandwidth of the remote entanglement generation. These ELUs are connected thorugh an OXC switch as shown in Fig.~\ref{MUSIQC}. (b) For QLA, each logic unit is made up of 49 physical qubits hosting four logical qubits and necessary ancilla qubits. (c) A logic block is six such logic units embedded in communication units. Communication units are square arrangements of $7 \times 7$ qubits, and eight such units fully surround the logic unit.}
\label{MUSIQCExample1}
\end{figure}

In our example, we assume multiplexities $m_p = 2$ and $m_T = 10$ that require 100 qubits ($= 3 \times 7 + 3\times 4 + 3 \times 2 \times 10$) and 12 parallel operations per ELU as shown in Fig. \ref{MUSIQCExample1}a. This choice adequately speeds up the communication time between ELUs to balance out other operation times in the hardware. Multiple ELUs are connected by an optical switch to complete the MUSIQC hardware (Fig. \ref{MUSIQC}b). With these resources, an efficient implementation of QCLA circuit can realized by executing all necessary logic gates in parallel. Under these circumstances, the depth of the $n$-bit in-place adder circuit is given by~\cite{DraperQIC2006}

\begin{equation}
\lfloor \log_2 n \rfloor + \lfloor \log_2 (n-1) \rfloor + \lfloor \log_2 \frac{n}{3} \rfloor +  \lfloor \log_2 \frac{n-1}{3} \rfloor +14,
\label{QCLADepth}
\end{equation}
for sufficiently large $n$ ($n>6$) where $\lfloor x \rfloor$ denotes the largest integer not greater than $x$. Out of these, two time steps contain $X$ gates, four contain $CNOT$ gates, and the rest contain Toffoli gates which dominate the execution time of the circuit. We assume an error correction step is performed on all qubits after each time step, by measuring all stabilizers of the Steane code and making necessary corrections based on the measurement outcome.

Once the basic operational primitives outlined in the previous section are modeled at the first level of code concatenation, we can construct all of these primitives at the second level of concatenation using the primitives at the first level. We can recursively construct the primitives at higher levels of code concatenation. Since the cost of remote CNOT gates between ELUs are independent of the distance between them, recursive estimation of circuit execution at higher levels of code concatenation is straightforward on MUSIQC hardware.

\subsection{QLA Implementation}
We consider a concrete layout of a QLA device optimized for $n$-bit adder with one level of Steane [[7,1,3]] encoding, which can be used to construct circuits at higher levels of code concatenation. In order to implement the fault-tolerant Toffoli gate described in Fig. \ref{FTToffoli}, one should assemble four logical qubits into a single tight unit, as we did for the ELUs in the MUSIQC architecture. In the QLA implementation, a ``Logic Unit (LU)'' consists of a square of 49 ($=7 \times 7$) qubits, where a block of 12 ($=3 \times 4$) qubits form a logical qubit with 7 physical qubits and 5 ancilla qubits (Fig. \ref{MUSIQCExample1}b). Just like in the MUSIQC example, $6n$ logical qubits placed on $1.5n$ LUs are necessary for adding two $n$ bit numbers. Therefore, we organize six LUs into a logical block (LB), capable of adding two 4 bit numbers. Each LU in the LB is surrounded by eight blocks of $7 \times 7$ communication units dedicated for distributing entanglement using the quantum repeater protocol (Fig. \ref{MUSIQCExample1}c). We assume that the communication of the qubits within each LU is ``free'', and do not consider the time it takes for such communication. This simplified assumption is justified as the communication time between LUs utilizing the qubits in the communication units dominate the computation time, and therefore does not change the qualitative conclusion of this estimate.

Similar to the MUSIQC hardware example, a Toffoli gate execution involves the preparation of the state $\ket{\phi_+}_L$ state in an ``empty'' LU, then teleporting three qubits onto this LU to complete the gate operation. The execution time of the Toffoli gate therefore is comprised of the time it takes to prepare the $\ket{\phi_+}_L$ state, the time it takes to distribute entanglement between adequate pairs of LUs, and then utilizing the distributed entanglement to teleport the gate operation. Among these, the distribution time for the entanglement is a function of the distance between the two LUs involved, while the other two are independent of the distance.

QCLA circuit involves various stages of Toffoli gates characterized by the ``distance'' between qubits that goes as $2^t$, where $1\leq t \leq \lfloor \log_2 n \rfloor$~\cite{DraperQIC2006}. In a 2D layout as considered in Fig. \ref{MUSIQCExample1}c, the linear distance between these two qubits goes as $\sqrt{2^t}$, in units of the number of communication units that the entanglement must be generated over. A slightly more careful analysis shows that the linear distance is approximately given by $d(t) \approx 3\cdot2^{t/2} +1$ when $t$ is even, and $d(t) \approx 2^{(t+1)/2}+1$ when $t$ is odd. Since each communication unit has 7 qubits along a length, the actual teleportation distance is $L(t) = 7d(t)$ in units of the length of ion chain. The nested entanglement swapping protocol can create entanglement between the two end ions in $\lfloor \log_2 L(t) \rfloor$ time steps, where each time step consists of one $CNOT$ gate, two single qubit gates, and one qubit measurement process. Using the expression for $d(t)$, we approximate $\log_2 L(t) \approx t/2+4$ for both even and odd $t$, without loss of much accuracy. Unlike in the case of MUSIQC, the entanglement generation time is now dependent on the distance between the qubits (although only in a logarithmic way), and the resulting time steps needed for entanglement distribution within the QCLA is (approximately) given by

\begin{widetext}
\begin{equation}
\lfloor \log_2 n \rfloor (\lfloor \log_2 n \rfloor\!+\!17)/4 \!+\! \lfloor \log_2 (n-1) \rfloor (\lfloor \log_2 (n-1) \rfloor\!+\!17)/4 \! \nonumber \\
+\! \lfloor \log_2 \frac{n}{3} \rfloor (\lfloor \log_2 \frac{n}{3} \rfloor\!+\!17)/4 \!+\!  \lfloor \log_2 \frac{n\!-\!1}{3} \rfloor ( \lfloor \log_2 \frac{n\!-\!1}{3} \rfloor\!+\!17)/4.
\label{QCLADepthQLA}
\end{equation}
\end{widetext}
It should be noted that in order to achieve this logarithmic time, one has to have the ability to perform two qubit gate between every pair of qubits in the entire communication units in parallel. The addition of two $n$ qubit numbers require $n/4$ LBs. Since each LB has 18 communication units, there are a total of $7 \times 7 \times 18 =882$ communication qubits in a LB. The number of parallel operations necessary is therefore 441 simultaneous $CNOT$ operations per LB, or $441n/4 \approx 110n$ parallel operations for $n$-bit QCLA. The number of $X$, $CNOT$ and Toffoli gates that have to be performed remains identical to the MUSIQC case since we are executing identical circuit. We assume that the error correction is performed after every logic gate, but the entanglement distribution process has high enough fidelity so that no further distillation process is necessary.

Similar to MUSIQC case, one can generate basic operational primitives at higher levels of code concatenation in the QLA model. Unlike the first encoding level, one may not have to explicitly provide communication channels for the second level of code concatenation if the quality of the distributed entanglement is sufficiently high so that neither entanglement purification~\cite{DurPRA1999} nor error correction of the entangled pairs~\cite{JiangPRA2009} is not needed. This type of ``inter-level optimization'' can be justified because the remote interaction between two logical qubits at second level of code concatenation occurs very rarely, and the communication units at the first level can be used to accommodate this communication at higher level without significant time overhead. If dedicated communication qubits were provided in addition, these qubits might sit idle most of the time leading to inefficient use of the qubit resources. The number of physical qubits therefore scale much more favorably at higher levels of code concatenation than in the first level in the QLA architecture. The distance-dependent gate operation at higher levels of code concatenation is somewhat difficult to predict accurately, but the logarithmic scaling of communication time allows effective estimation of the gate operation time with only small errors.

\subsection{Results and Comparison}

\begin{figure}
\includegraphics[width=1.0\linewidth]{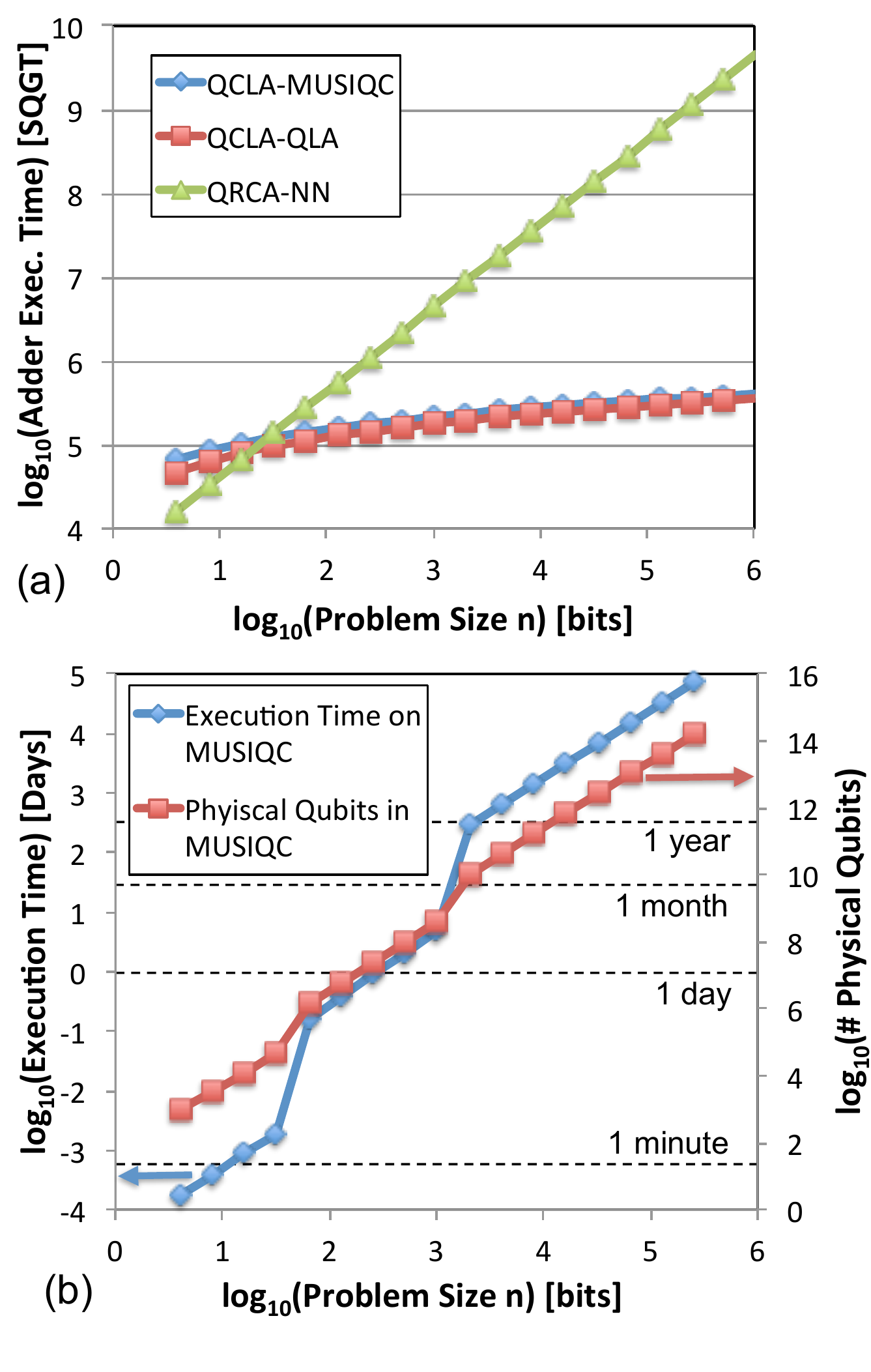}
\caption{(Color online) (a) Execution time comparison of quantum ripple-carry adder (QRCA) on a nearest-neighbor architecture (green triangles), and quantum carry-lookahead adder (QCLA) on QLA (red squares) and MUSIQC (blue diamonds) architectures, as a function of the problem size $n$. All three circuits considered are implemented fault-tolerantly, using one level of Steane [[7,1,3]] code. The execution time is measured in units of single qubit gate time (SQGT), assumed to be $1\mu$sec in our model. (b) Execution time (blue diamonds, left axis) and number of required physical qubits (red squares, right axis) of running fault-tolerant modular exponentiation circuit, representative of executing the Shor algorithm.}
\label{AdderPerformance}
\end{figure}

 Figure \ref{AdderPerformance}a  and Table \ref{ResultSummary} summarize the resource requirements and performance of the QCLA circuit on MUSIQC and QLA architecture, as well as QRCA circuit on a nearest neighbor (NN) quantum hardware, where multi-qubit gates can only operate on qubits sitting right next to one another. Although the QLA architecture considered in this example is also a NN hardware, presence of the dedicated communication units (quantum bus) allows remote gate operation with the execution time that depends only logarithmically on the distance between qubits, enabling fast execution of the QCLA. The cost in resources, however, is significant: realization of efficient communication channel requires $\sim 3$ times as many physical qubits as used for storing and manipulating the qubits in the first level of encoding, and requires a large number of parallel operations and the necessary control hardware to run them. The execution time can be fast compared to the MUSIQC architecture, which is hampered by the probabilistic nature of the photonic network in establishing the entanglement. We have dedicated substantial resources in MUSIQC to speed up the entanglement generation time as described in the previous section. Although MUSIQC architecture will take $\sim 15-30 \%$ more time to execute the adder circuit, the resources it requires to operate the same task is only about 13\% of that required in the QLA architecture. In both cases, we note the importance of moving qubits between different parts of a large quantum computer. The speed advantage in adder circuits translate directly to faster execution of Shor algorithm, so we adopted QCLA for further analysis.

\begin{table}
\caption{Summary of the resource estimation and execution times of various adders in MUSIQC and QLA architecture.}
\begin{ruledtabular}
\begin{tabular}{c | c c c} 
Performance & QCLA on & QCLA on  & QRCA on\\
Metrics & MUSIQC & QLA  & NN\\ \hline
Physical Qubits & 150$n$ & 1,176$n$  & 20($n$+1)  \\ 
\# Parallel Operations & 18$n$ & 110$n$ & 8$n+43$\\
Logical Toffoli ($\mu$s) & 3,250 & 2,327\footnote{Does not include entanglement distribution time} & 2,159 \\
128-bit addition & 0.16 s & 0.13 s & 0.56 s \\
1,024-bit addition & 0.22 s & 0.18 s & 4.5 s\\
16,384-bit addition & 0.29 s & 0.25 s & 72 s\\
\end{tabular}
\end{ruledtabular}
\label{ResultSummary}
\end {table}

Once the execution time and resource requirements are identified for the adder circuit, one can adopt the analyses provided in Ref. \cite{VanMeterPRA2005} to estimate the performance metrics of running Shor algorithm. The execution time and total number of physical qubits necessary to run Shor algorithm depends strongly on the level of code concatenation required to successfully obtain the correct answer. We first estimate the number of logical qubits ($Q$) and the total number of logic gate operations ($K$) required to complete the Shor algorithm of a given size, to obtain the product $KQ$. In order to obtain correct results with a probability of order unity, the individual error rate corresponding to one logic gate operation must be on the order of $1/KQ$ \cite{MetodiMicro2005}. From this consideration, we determine the level of code concatenation to be used.  Table \ref{ShorSummary} summarizes the comparison on the number of physical qubits and the execution time of running Shor algorithm on MUSIQC and QLA architectures for factoring 32, 512 and 4,096 bit numbers \cite{ClarkPRA2009}.

\begin{table}
\caption{Estimated execution time and physical qubits necessary to complete Shor algorithm of a given size. The numbers on top (bottom) correspond to MUSIQC (QLA) architecture.}
\begin{ruledtabular}
\begin{tabular}{c | c c c c} 
Performance & & $n=32$ & $n=512$  & $n=4,096$\\
Metrics &  & &  & \\ \hline
Code Level & & 1 & 2  & 3  \\ \hline
\# Physical  & MUSIQC & $4.7\times10^4$ & $9.2\times10^7$ & $4.1\times10^{10}$\\
Qubits & QLA & $3.7\times10^5$ & $7.2\times10^8$ & $3.2\times10^{11}$\\ \hline
Execution  & MUSIQC & 2.5 min & 2.1 days & 650 days \\
Time & QLA & 2.2 min & 1.5 days & 520 days\\
\end{tabular}
\end{ruledtabular}
\label{ShorSummary}
\end {table}

Figure \ref{AdderPerformance}b shows the execution time (in days) and the total number of necessary physical qubits for completing the modular exponentiation circuit on a MUSIQC hardware, which is a good representation of running the Shor algorithm.  The discrete jumps in the resource estimate correspond to addition of another level of code concatenation, necessary for maintaining the error rates low enough to obtain a correct result as the problem size increases. Using 2 levels of concatenated Steane code, we expect to be able to factor a 128-bit integer in less than 10 hours, with less than $6\times10^6$ physical qubits in the MUSIQC system. The execution time on QLA architecture is comparable to that on MUSIQC architecture (within ~20\%), but the number of required physical qubits is higher by about a factor of 10. Furthermore, the total size of the single ELU necessary to implement the QLA architecture grows very quickly (over $4.5\times10^7$ physical qubits for machine that can factor a 128-bit number), while the ELU size in MUSIQC architecture is fixed at moderate numbers ($\approx 58,000$ ELUs with 100 qubits per ELU). Therefore, although still daunting, the MUSIQC architecture substantially lowers the practical technological barrier in integration levels necessary for a large-scale quantum computer.

\section{Fault Tolerance of Probabilistic Photonic Gates}

Na\" ively, it would appear that the average entanglement creation time $\tau_E$ must be much smaller 
than the decoherence time scale $\tau_D$ to achieve fault tolerance, but we find that scalable fault-tolerant quantum computation is possible for {\em{any}} ratio $\tau_E/\tau_D$, even in the presence of additional gate errors.  While large values of $\tau_E/\tau_D$ would lead to impractical levels of overhead in qubits and time (similar to the case of conventional quantum fault tolerance near threshold error levels \cite{Knill05}), this result is still remarkable and indicates that fault tolerance is always possible in the MUSIQC architecture. In this section, we provide a complete description of the strategies used to secure fault tolerance in MUSIQC architecture.

\subsection{Analysis of fault-tolerance for fast entangling gates}
\label{fastgates}
First, we consider the case where $\tau_E/\tau_D \ll 1$, where fault tolerant coding is more practical. When each ELU is large enough to accommodate logical qubits encoded with a conventional error correcting code, one can implement full fault-tolerant procedure within an ELU as in the example presented in the previous section. When the ELUs are too small to fit the logical qubits, fault-tolerance can be achieved by mapping to three-dimensional (3D) cluster states, a known approach for supporting fault-tolerant universal quantum computation \cite{RHG06}. This type of encoding is well-matched to the MUSIQC architecture, because the small degree of their interaction graph leads to small ELUs.

\begin{figure}
\begin{center}
\includegraphics[width=1.0\linewidth]{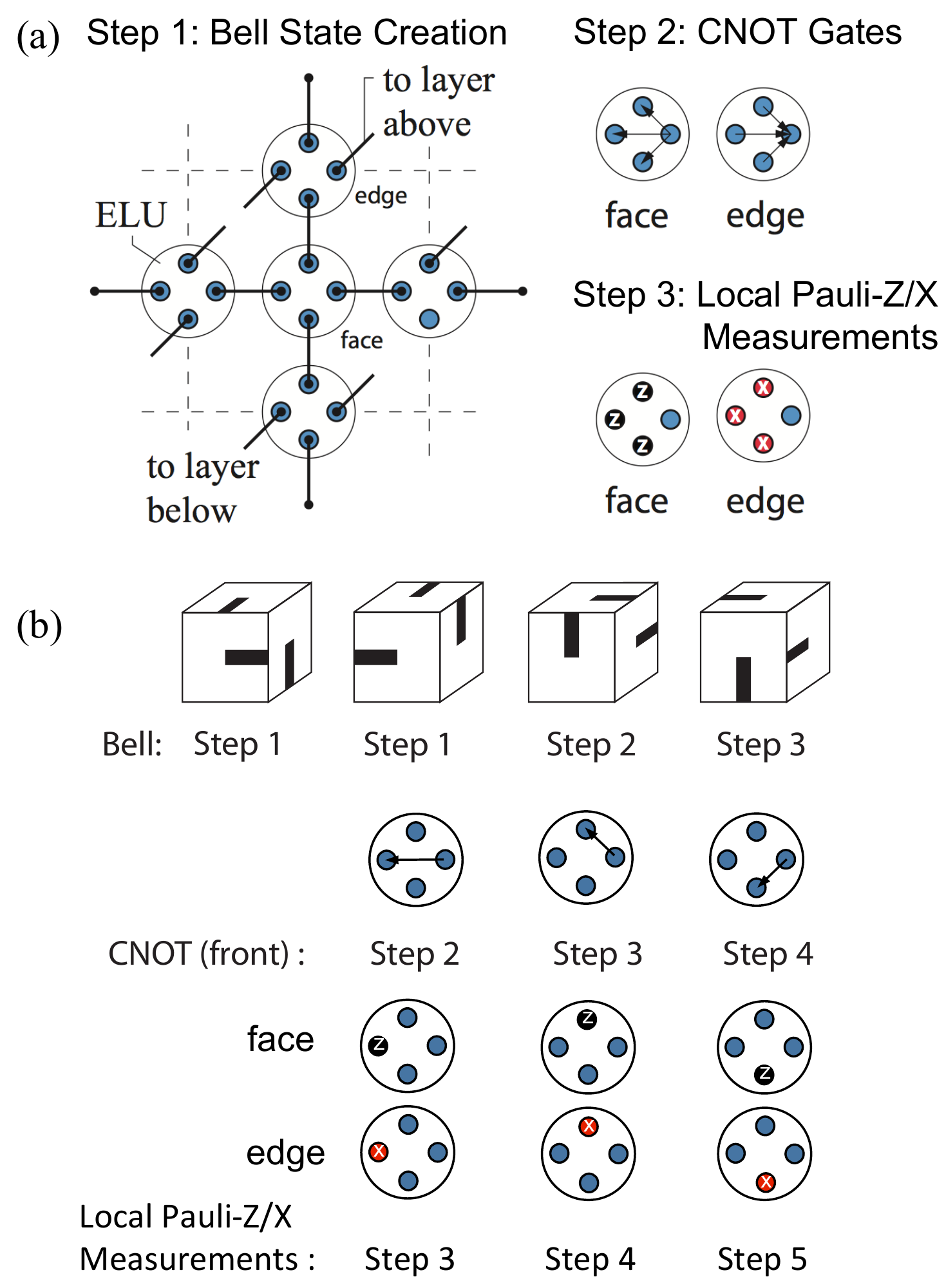}
 \caption{\label{CluCr} (a) Three steps of creating a 3D cluster state in the MUSIQC architecture, for fast entangling gates. Step 1: Creation of Bell pairs between different ELUs, all in parallel. Step 2: CNOT gates (head of arrow: target qubit, tail of arrow: control qubit). Step 3: Measuring of 3 out of 4 qubits per ELU. If the ELU represents a face (edge) qubit in the underlying lattice, the measurements are in the $Z$- ($X$-) basis. The resulting state is a 3D cluster state, up to Hadamard gates on the edge-qubits. (b) Schedule for the creation of a 3D cluster state in the MUSIQC architecture. Upper line: Schedule for Bell pair production between ELUs representing face and edge qubits. Lower line: Schedule for the CNOT gates within the ELUs corresponding to the front faces of the lattice cell. Schedules for the ELUs on other faces and on edges are similar.}
\end{center}
\end{figure}

{\em{Scheduling.}} For $\tau_E\ll\tau_D$, the 3D cluster state with qubits on the faces and edges of a three-dimensional lattice can be created using the procedure displayed in Fig.~\ref{CluCr}a. The procedure consists of three basic steps: (1) Creation of Bell states between different ELUs via the photonic link, (2) CNOT-gates within each ELU, and (3) local measurement of three out of four qubits in each ELU. As can be easily shown using standard stabilizer arguments, the resulting state is a 3D cluster state, up to local Hadamard gates on the edge qubits.

The operations can be scheduled such that (a) qubits are never idle, and (b) no qubit is acted upon by multiple gates (even commuting ones) at the same time. The latter is required in some proposals for realizing quantum gates with ion qubits. To this end, the schedule \cite{RHG06} for 3D cluster state generation is adapted to the MUSIQC architecture, and the three-step sequence shown in Fig.~\ref{CluCr}a is expanded into the five-step sequence shown in Fig.~\ref{CluCr}b. Through Steps 1 - 3 the Bell pairs across the ELUs are created. Through Steps 2 - 4 the CNOTs within each ELU are performed, and through Steps 3 - 5 three qubits in each ELU are measured. The sequence of operations is such that each of the three ancilla qubits in every ELU lives for only three time steps: initialization (to half of a Bell pair), CNOT, measurement. No qubit is ever idle in this protocol. 

What remains to complete the computation is the local measurement of the 3D cluster state \cite{RHG06}. All remaining measurements are performed in Step 5 of the above procedure. This works trivially for cluster qubits intended for topological error correction or the implementation of topologically protected encoded Clifford gates \cite{RHG07}, since these measurements require no adjustment of the measurement basis. To avoid delay in the measurement of qubits for the implementation of non-Clifford gates, it is necessary to break the 3D cluster states into overlapping slabs of bounded thickness \cite{RHG06}.  

{\em{Fault-tolerance threshold.}} We assume the following error model. (1) Every gate operation, i.e. preparation and measurement of individual qubits, gates within an ELU, and Bell pair creation between different ELUs, can all be achieved within a clock cycle of duration $T$.  An erroneous one-qubit (two-qubit) gate is modeled by the perfect gate followed by a partially depolarizing one-qubit (two qubit) channel. In the one-qubit channel, $X$, $Y$, and $Z$ errors each occur with probability $\epsilon/3$. In the two-qubit channel, each of the 15 possible errors $X_1$,$X_2$,$X_1X_2$, .. ,$Z_1Z_2$ occurs with a probability of $\epsilon/15$. All gates have the same error $\epsilon$.  (2) In addition, the effect of decoherence per time step $T$ is described by local probabilistic Pauli errors $X$, $Y$, $Z$, each happening with a probability $T/3\tau_D$.

\begin{figure}
\begin{center}
\includegraphics[width=0.92\linewidth]{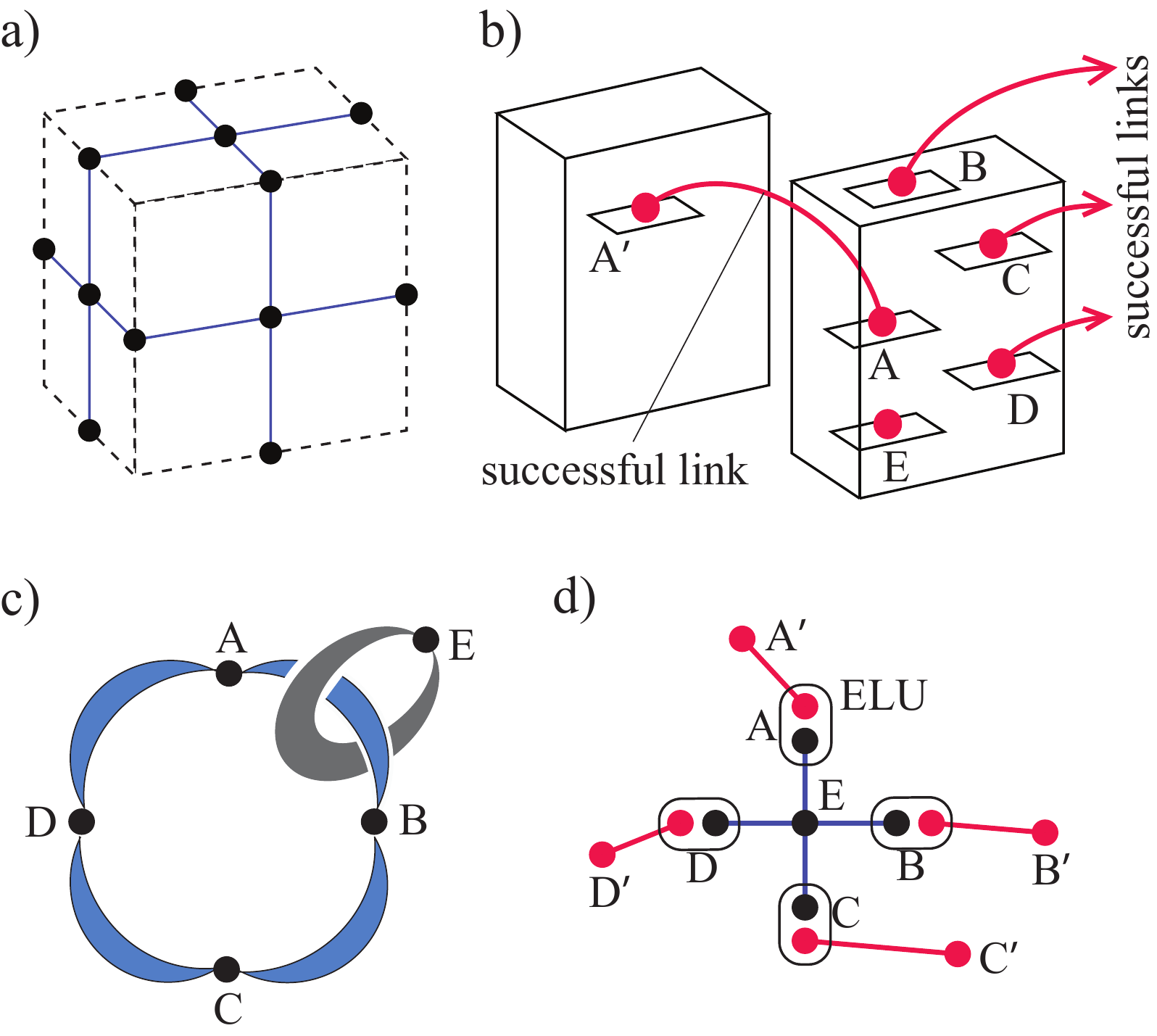}
\caption{\label{HC2} Hypercell construction II. (a) Lattice cell of a three-dimensional four-valent cluster state. The dashed lines represent the edges of the elementary cell and the full lines represent the edges of the connectivity graph. The three-dimensional cluster state is obtained by repeating this elementary cell in all three spatial directions. (b) Creating probabilistic links between several 3D cluster states. (c) Reduction of a 3D cluster state to a 5-qubit graph state, via Pauli measurements. The shaded regions represent measurements of $Z$, the blank regions represent measurements of $X$. The qubits represented as black dots remain unmeasured. For details, see \cite{RHG06}. (d) Linking graph states by Bell measurements in the remaining ELUs. Four-valent, 3D cluster states of arbitrary size can be created.}
\end{center}
\end{figure}

A criterion for the error threshold of measurement-based quantum computation with cluster states that has been established numerically for a variety of error models is 
\begin{equation}
\label{Kcrit}
\langle K_{\partial q}\rangle(\{\mbox{error parameters}\}) = 0.70. 
\end{equation}
Therein, $K_{\partial q}$ is a cluster state stabilizer operator associated with the boundary of a single volume $q$, consisting of six faces. Let $f$ be a face of the three-dimensional cluster, and $K_f = \sigma_{x}^{(f)}\bigotimes_{e \in \partial f}\sigma_{z}^{(e)}$ as shown in Fig.~\ref{HC2}a. Then, $K_{\partial q} = \prod_{f \in \partial q}K_f = \bigotimes_{f \in \partial q}\sigma_x^{(f)}$. Furthermore, for the above criterion to apply, all errors--for preparation of local states, local and entangling unitaries, and measurement--are propagated forward or backward in time, to solely affect the 3D cluster state. 

The above criterion applies for a phenomenological error-model with local memory error and measurement error (the threshold error probability per memory step and measurement is 2.9\% \cite{HWang}), for a gate-based error model (the threshold error probability per gate is 0.67\% \cite{RHG06}), and further error models with only low-order correlated error. Specifically, the criterion (\ref{Kcrit}) has numerically been tested for cluster state creation procedures with varying relative strength of local vs 2-local gate error \cite{RHG06}, with excellent agreement. In all cases, the error-correction was performed using Edmonds' perfect matching algorithm.

The detailed procedure for calculating the error probability of the stabilizer measurement process for the 3D cluster state is provided in Appendix~\ref{ErrorRateCalc}. In combination with Criterion~(\ref{Kcrit}), we obtain the threshold condition 

\begin{equation}
  \label{ThrC}
  \epsilon + \frac{55}{32} \frac{T}{\tau_D} < 2.9\times 10^{-3}.
\end{equation} 

{\em{Overhead.}} The operational cost of creating a 3D cluster state and then locally measuring it for the purpose of computation is 24 gates per elementary cell in the standard setting, and 54 gates per elementary cell in MUSIQC. Here the elementary cell of a 3D four-valent cluster state is shown in Fig.~\ref{HC2}b.  The overhead of the MUSIQC architecture over fault-tolerant cluster state computation is thus constant. The operational overhead for fault-tolerance in the latter is poly-logarithmic \cite{RHG06}, as described in detail in Ref. \cite{RHG07}.\medskip

\subsection{Analysis of fault-tolerance for slow entangling gates}
The above construction fails for $\tau_E/\tau_D\geq 1$, where decoherence occurs while waiting for Bell-pair entanglement. However, scalable fault-tolerant computing can still be achieved in the MUSIQC architecture for {\em any} ratio $\tau_E/\tau_D$, even for ELUs of only 3 qubits. Compared to the case of $\tau_E\ll \tau_D$, the operational cost of fault-tolerance is increased by a factor that depends strongly on $\tau_E/\tau_D$ but is {\em independent} of the size of the computation. Thus, while quantum computation becomes more costly when $\tau_E\geq \tau_D$, it remains scalable. This surprising result shows that there is no hard threshold for the ratio $\tau_E/\tau_D$, and opens up the possibility for efficient fault-tolerant constructions with slow entangling gates. Here we show that scalable quantum computation can be achieved for arbitrarily slow entangling gates.

The main idea is to construct a ``hypercell'' out of several ELUs. A hypercell has the same storage capacity for quantum information as a single ELU, but with the ability to become (close to) deterministically entangled with four other hypercells. Fault-tolerant universal quantum computation can then be achieved by mapping to a 4-valent, three-dimensional cluster state \cite{RHG06}. First, we show that arbitrarily large ratios $\tau_E/\tau_D$ can be tolerated in the limiting case where the gate error rate $\epsilon = 0$ (Construction I). Then, we show how to tolerate arbitrarily large ratios $\tau_E/\tau_D$ with finite gate errors $\epsilon > 0$ (Construction II).  

\begin{figure}
\begin{center}
\includegraphics[width=1.0\linewidth]{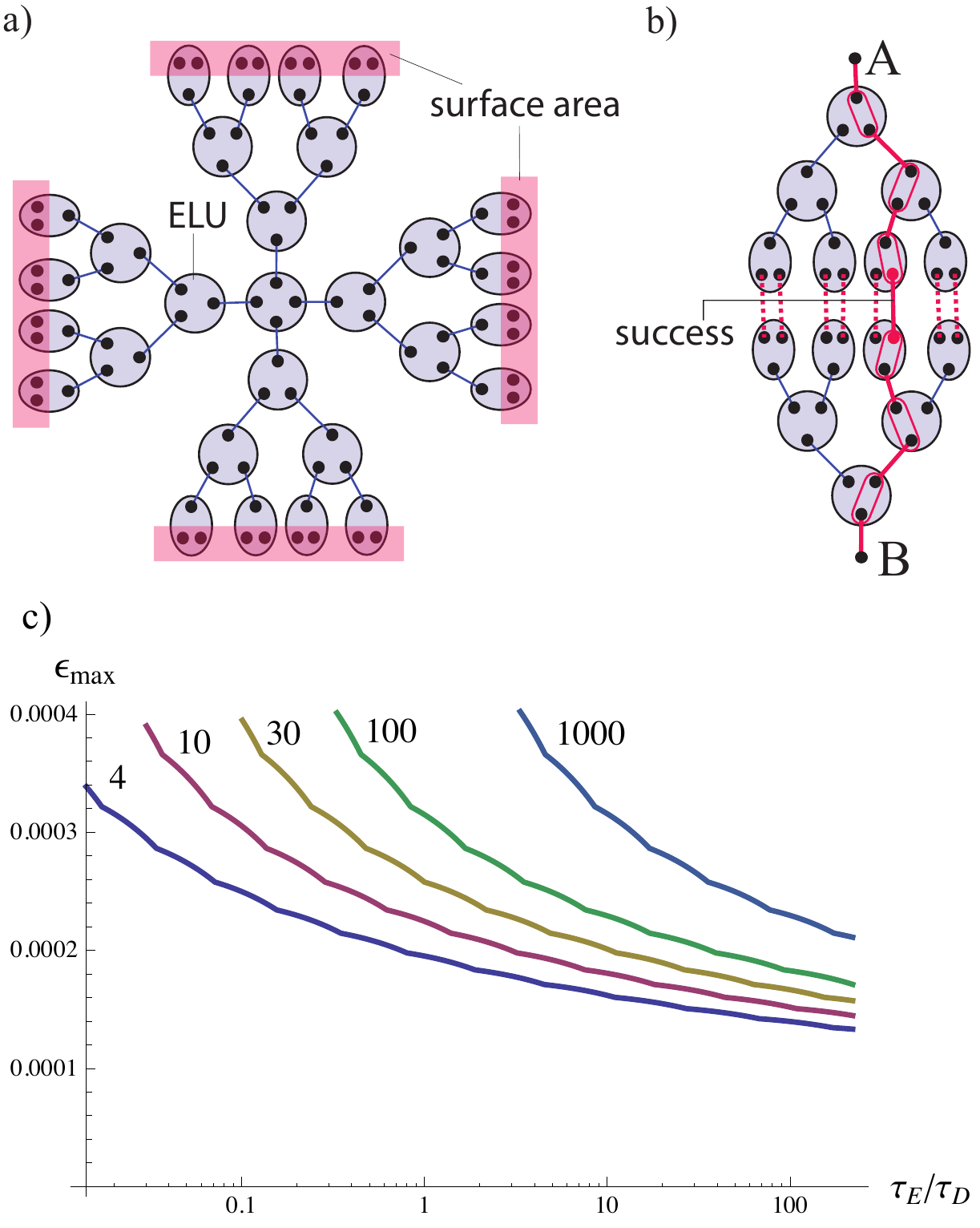}
\caption{\label{tree} (Color online) Hypercell construction I. (a) Snowflake design of Refs.~\cite{LiPRL2010,FujiiPRL2010}. (b) Connecting two hypercells. If the surface area is large, with high probability one or more Bell pairs are created between the surface areas via the photonic link. By Bell measurements within individual ELUs (indicated by ovals) one such Bell pair is teleported to the roots $A$ and $B$. c) Boundary of the fault-tolerance region for gate error $\epsilon$ and ratio $\tau_E/\tau_D$, for various ELU sizes. The threshold for the gate error $\epsilon$ depends only weakly on $\tau_E/\tau_D$.}
\end{center}
\end{figure} 

{\em{Hypercell Construction I}} is based on the snowflake design \cite{LiPRL2010,FujiiPRL2010}, as shown in Fig.~\ref{tree}a. The difference is that in the present case, each node in the connectivity tree represents an entire ELU, not a single qubit as in Refs.~\cite{LiPRL2010,FujiiPRL2010}. At the root of the tree is an ELU that contains the qubit used in the computation, while multiple layers of bifurcating branches lead to a  large ``surface area'' with many ports from which to attempt entanglement generation between two trees. Once a Bell pair is created, it can be converted to a Bell pair between the root qubits $A$ and $B$ via teleportation as shown in Fig.~\ref{tree}b.   

The links (each representing a Bell pair) within a snowflake structure are created probabilistically, each with a probability $p$ of heralded success. The success probability of each hypercell is small, but if the surface area between two neighboring hypercells is large enough, the probability of creating a Bell pair between them via a probabilistic photonic link approaches unity.  Thus, the cost of entangling an entire grid of hypercells is linear in the size of the computation, as opposed to the exponential dependence that would be expected if the hypercells could not be entangled deterministically.  Correspondingly, the operational cost of creating a hypercell is large, but the cost of linking this qubit into the grid is independent of the size of the computation.  The hypercell offers a qubit which can be near-deterministically entangled with a constant number of other qubits {\em{on demand}}.  A quantum computer made up of such hypercells can create a four-valent, 3D cluster state with few missing qubits, and is thus fault-tolerant  \cite{RHG06}, \cite{Barrett2}, \footnote{If ELUs of size $N_q=3$ are used, resulting in hypercells of valency 3, then two such hypercells can be combined into one of valency 4.}. Hypercells can readily be implemented in the modular ion trap quantum computer since the probability of entanglement generation does not depend on the physical distance between the ELUs.

We call the part of the hypercell needed to connect to a neighboring hypercell a ``tree''. For ELUs of coordination number 3, the number $m$ of ports that are available to connect two hypercells is twice the number of ELUs in the top layer of the tree. The probability for all $m$ attempts to generate entanglement between two trees to fail is $P_{\text{fail}}=(1-p)^m\approx \exp(-mp)$. (In practice, we will allow a constant probability of failure which is tolerable in 3D cluster states \cite{Barrett2}.) 
In addition, the number of ELUs in the top layer is $2^{\#\, \text{layers}}$, and the path length $l$ (number of Bell pairs between the roots) is $l=2 \log_2m + 1$. Combining the above, we find that $l = 2 \log_2 \frac{c}{p}\, + 1$,
for $c=-\ln P_{\text{fail}}$. For simplification we assume that the time $t$ for attempting entanglement generation is the same when creating the trees and when connecting the trees. Then, $p = t/\tau_E$ in both cases. From the beginning of the creation of the trees to completion of entangling two trees, a time $2t$ has passed. The Bell pairs within the trees have been around, on average, for a time $3t/2$, and the Bell pairs between the two trees for an average time of $t/2$. If overall error probabilities remain small, the total probability of error for creating a Bell pair is proportional to $l$. The memory error alone is
\begin{equation}
  \label{Emem}
  \epsilon_{\text{mem}} = \frac{t}{\tau_D}\left[3 \log_2\left(c\frac{\tau_E}{t}\right) + \frac{1}{2}\right]. 
\end{equation}
This function is monotonically increasing with $t$, and $\epsilon_{\text{mem}}(t=0)=0$. The task now is to suppress the memory error rate $\epsilon_{\text{mem}}$ below the error threshold $\epsilon_{\text{crit}}$ that applies to fault-tolerant quantum computation with 3D cluster states. From Eq.~(\ref{Kcrit}) we know that $\epsilon_{\text{crit}}>0$. 

From Eq.~(\ref{Emem}) we find that, for any ratio $\tau_E/\tau_D$, we can make $t$ small enough such that $\epsilon_{\text{mem}}<\epsilon_{\text{crit}}$. The operational cost for creating a hypercell with sufficiently many ports is
$O(\text{hypercell}) \sim \left(\frac{1}{p}\right)^{\frac{9/2\,c}{p}}$.
This cost is high for small $p=t/\tau_E$, but independent of the size of the computation. Thus, whenever decoherence on waiting qubits is the only source of error, scalable fault-tolerant QC is possible for arbitrarily slow entangling gates. \medskip

We now discuss how the above Hypercell Construction I fares in the presence of additional gate error $\epsilon$. We model every noisy one-(two-)qubit operation by the perfect operation followed by a $SU(2)$- ($SU(4)$-) invariant partial depolarizing channel with strength $\epsilon$, same as that used in Section~\ref{fastgates}. If $\epsilon >0$ then every entanglement swap adds error to the computation. We must swap entanglement in every ELU on the path between the roots $A$ and $B$, and because there are $2\log_2 m$ of them ($m \geq 2$), for $\epsilon \ll 1$ the total error is
\begin{equation}
  \label{Et}
  \epsilon_{\text{total}} = \frac{t}{\tau_D}\left[3 \log_2\left(c\frac{\tau_E}{t}\right) + \frac{1}{2}\right] + 2 \epsilon \log_2\left(c\frac{\tau_E}{t}\right). 
\end{equation}
Now it is no longer true that for any choice of $\tau_E/\tau_D$ we can realize $\epsilon_{\text{crit}}> \epsilon_{\text{total}}$. A non-vanishing gate error sets an upper limit to the tree depth, because the accumulated gate error is proportional to the tree depth (Fig.~\ref{tree}b). This implies an upper bound on the size of the top layer of the tree, which further implies a lower bound on the time $t$ needed to attempt entangling the two trees (see Eq.~(\ref{C1}) below) and thus a lower bound on the memory error caused by decoherence during the time interval $t$. The accumulated memory error alone may be above or below the error threshold, depending on the ratio $\tau_E/\tau_D$. 

In more detail, suppose that $\epsilon_{\text{crit}}> \epsilon_{\text{total}}$ holds.  Considering only gate errors, $\epsilon_{\text{crit}} >  2 \epsilon \log_2\left(c\frac{\tau_E}{t}\right)$, and hence, 
\begin{equation}
\label{C1}
t > c \tau_E 2^{-\frac{\epsilon_{\text{crit}}}{2\epsilon}}.
\end{equation}
Now, recalling that $c\frac{\tau_E}{t} = m \geq 2$, with Eq.~(\ref{Et}) we find that $\epsilon_{\text{crit}} > 3t/\tau_D+ 2\epsilon$, or
\begin{equation}
  \label{C2}
  t < \frac{1}{3} (\epsilon_{\text{crit}}-2\epsilon) \tau_D.
\end{equation}
The two conditions Eq.~(\ref{C1}) and (\ref{C2}) can be simultaneously obeyed only if
\begin{equation}
\label{C3}
  \frac{\tau_E}{\tau_D} < \frac{\epsilon_{\text{crit}} - 2\epsilon}{3c} 2^{\frac{\epsilon_{\text{crit}}}{2\epsilon}}
\end{equation}
We see that there is now an upper bound to the ratio $\tau_E/\tau_D$. Eq.~(\ref{C3}) is a necessary but not sufficient condition for fault-tolerant quantum computation using the hypercells of Fig.~\ref{tree}b. 

We have numerically simulated the process of constructing these hypercells for various values of the decoherence parameters $\epsilon$ and $\tau_E/\tau_D$. The boundary of the fault-tolerance region in the $\tau_E/\tau_D,\epsilon$-plane is shown in Fig.~\ref{tree}c. In the above, for simplicity, we have considered hypercells in which all constituent ELUs are entangled in a single timestep $t$.  However, there are various possible refinements. (1) The computational overhead can be significantly decreased by creating the hypercell in stages, starting with the leaves of the trees and iteratively combining them to create the next layers \cite{LiPRL2010}. (2) Using numerical simulations it was found that if each of the 4 trees making up a hypercell has coordination number 4 or 5 rather then 3 (i.e., a ternary tree instead of a binary tree), the overhead can be further reduced.  These optimizations were used to produce Figure~\ref{tree}c.\medskip

{\em{Hypercell Construction II}} allows fault-tolerance for finite gate errors $\epsilon>0$. In Construction I, the accumulated error for creating a Bell pair between the roots $A$ and $B$ is linear in the path length $l$ between $A$ and $B$.  This limits the path length $l$, and thereby the surface area of the hypercell. This limitation can be overcome by invoking three-dimensional (3D) cluster states already at the level of creating the hypercell. 3D cluster states have an intrinsic capability for fault-tolerance \cite{RHG06} related to quantum error correction with surface codes \cite{Kit97,DKLP}. For Hypercell Construction II, we employ a 3D cluster state nested within another 3D cluster state. Therein, the ``outer'' cluster state is created near-deterministically from the hypercells. Its purpose is to ensure fault-tolerance of the construction. The ``inner'' 3D cluster state is created probabilistically. Its purpose is to provide a means to connect distant qubits in such a way that the error of the operation does not grow with distance. Specifically, if the local error level is below the threshold for error-correction with 3D cluster states, the error of (quasi-) deterministically creating a Bell pair between two root qubits $A$ and $B$ in distinct 3D cluster states is independent of the path length between $A$ and $B$. 

The construction is as follows. We start from a three-dimensional grid with ELUs on the edges and on the faces. Each ELU contains four qubits and can be linked to four neighboring ELUs. Such a grid of ELUs (of suitable size) is used to probabilistically create a 4-valent cluster state by probabilistic generation of Bell pairs between the ELUs, post-selection and local operations within the ELUs. 

After such cluster states have been successfully created, in each ELU three qubits are freed up, and can now be used for near-deterministic links between different 3D cluster states, as shown in Fig.~\ref{HC2}b. After 4 probabilistic links to other clusters have succeeded (the size of the cluster states is chosen such that this is a likely event), the cluster state is transformed into a star-shaped graph state via $X$ and $Z$ measurements (Fig.~\ref{HC2}c). This graph state contains 5 qubits, shared between the 4 ELUs at which the successful links start, and an additional ELU. Due to the topological error-correction capability of 3D cluster states, the conversion from the 3D cluster state to the star-shaped graph state is fault-tolerant \cite{RHG06}. By further measurement in the ELUs, the graph states created in different hypercells can now be linked, e.g. to form again a 4-valent 3D cluster state which is a resource for fault-tolerant quantum computation \cite{RHG06}, as shown in Fig.~\ref{HC2}d. This final linking step is prone to error. However, the error level is independent of the size of the hypercell, which was not the case for Hypercell Construction I.

The only error sources remaining after error-correction in the 3D cluster stem from (i) the (two) ports per link, and (ii) the two root qubits $A$ and $B$, which are not protected topologically. The total error $\epsilon_{\text{total}}$ of a Bell pair created between $A$ and $B$ in this case is given by
$
  \epsilon_{\text{total}} = c_1 t/\tau_D + c_2 \,\epsilon, 
$
where $t$ is the time spent attempting Bell pair generation, and $c_1$ and $c_2$ are algebraic constants which do not depend on the time scales $\tau_E$ and $\tau_D$, and not on the distance between the root qubits $A$ and $B$. Then, if the threshold error rate $\epsilon_{\text{crit}}$ for fault-tolerance of the outer 3D cluster state is larger than $c_2\, \epsilon$, we can reach an overall error $\epsilon_{\text{total}}$ below the threshold value $\epsilon_{\text{crit}}$ by making $t$ sufficiently small. Smaller $t$ requires larger inner 3D cluster states, but does not limit the success probability for linking Construction II hypercells. Thus, fault-tolerance is possible for all ratios $\tau_E/\tau_D$, even in the presence of small gate errors.

\section{Outlook}

The success of silicon-based information processors in the past five decades hinged upon the scalability of integrated circuits (IC) technology characterized by Moore's law~\cite{MooreE1965}. IC technology integrated all the components necessary to construct a functional circuit, using the same conceptual approach over many orders of magnitude in integration levels. The hierarchical modular ion trap quantum computer architecture discussed here promises scalability, not only in the number of physical systems (trapped ions) that represent the qubits, but also in the entire control structure to manipulate each qubit at such integration levels. 

The technology necessary to realize each and every component of the MUSIQC architecture is either already available or within reach.
The recognition that ion traps can be mapped onto a two dimensional surface that can be fabricated using standard silicon microfabrication technologies~\cite{NISTsurface05,Kim05} has led to a rapid development in complex surface trap technology~\cite{Sandia,GTRI}. Present-day trap development exploits extensive electromagnetic simulation codes to design optimized trap structures and control voltages, allowing sufficient control and stability of ion positioning. Integration of optical components into such microfabricated traps will enable stronger interaction between the ions and photons for better photon collection and qubit detection~\cite{KimQIC2009} through the use of high numerical aperture optics or integration of an optical cavity with the ion trap~\cite{KimMaunzKimPRA2011}. Moreover, electro-optic and MEMS-based beam steering systems allows the addressing of individual atoms in a chain with tightly focused laser beams~\cite{SchmidtKaler2003,KnoernschildAPL2010} and an optical interconnect network can be constructed using large-scale all-optical crossconnect switches~\cite{KimPTL2003}.
While technical challenges such as the operation of narrowband (typically ultraviolet) lasers
or the presence of residual heating of ion motion \cite{NIST} still remain, they do not appear to be fundamental roadblocks to scalability.  Within the MUSIQC architecture we have access to a full suite of technologies to realize the ELU in a scalable manner, where the detailed parameters of the architecture such as the number of ions per ELU, the number of ELUs, or the number of photonic interfaces per ELU can be adapted to optimize performance of the quantum computer. 

\section{Acknowledgements}
We thank D. Bacon, M. Biercuk, B. Blinov, S. Flammia, D. L. Moehring, and R. E. Slusher for helpful discussions. This work was supported by the Intelligence Advanced Research Projects Activity, the Army Research Office MURI Program on Hybrid Quantum Optical Circuits, and the NSF Physics Frontier Center at JQI.  LMD acknowledges support by the NBRPC (973 Program) 2011CBA00300.

\appendix
\section{Universal Fault-Tolerant QC using Steane Code}
\label{UniversalFTGates}
We utilize the basic operational primitives of universal quantum computation using Steane [[7,1,3]] code \cite{SteanePRL1996} fully outlined in Ref. \cite{MikeAndIke}, summarized below.

\begin{enumerate}
\item The preparation of logical qubit $\ket{0}_L$ is performed by measuring the six stabilizers of the code using four-qubit cat state $\ket{cat}_4 \equiv (\ket{0000}+\ket{1111})/\sqrt{2}$, following the procedure that minimizes the use of ancilla qubits as outlined in Ref. \cite{DiVincenzo07}. The stabilizer measurement is performed up to three times to ensure that the error arising from the measurement process itself can be corrected. We perform a sequential measurement of the six stabilizers re-using the four ancilla qubits for each logical qubit, which reduces the number of physical qubits and parallel operations necessary for the state preparation at the expense of the execution time. Once all the stabilizers are measured, a three-qubit cat state is used to measure the logical $Z_L$ operator to finalize qubit initialization process. This procedure requires eleven physical qubits to complete preparation of logical qubit $\ket{0}_L$.

\item In Steane [[7,1,3]] code considered here, all operations in the Pauli group $\{ X_L, Y_L, Z_L \}$ and the Clifford group $\{H_L, S_L, CNOT_L \}$ can be performed transversally ({\em i.e.}, in a bit-wise fashion). We assume seven parallel operations are available, so that these logical operations can be executed in one time step corresponding to the single- or two-qubit operation. The transversal $CNOT_L$ considered here is between two qubits that are close by, so the operation can be performed locally without further need for qubit communication.

\begin{figure}[b]
\includegraphics[width=0.92\linewidth]{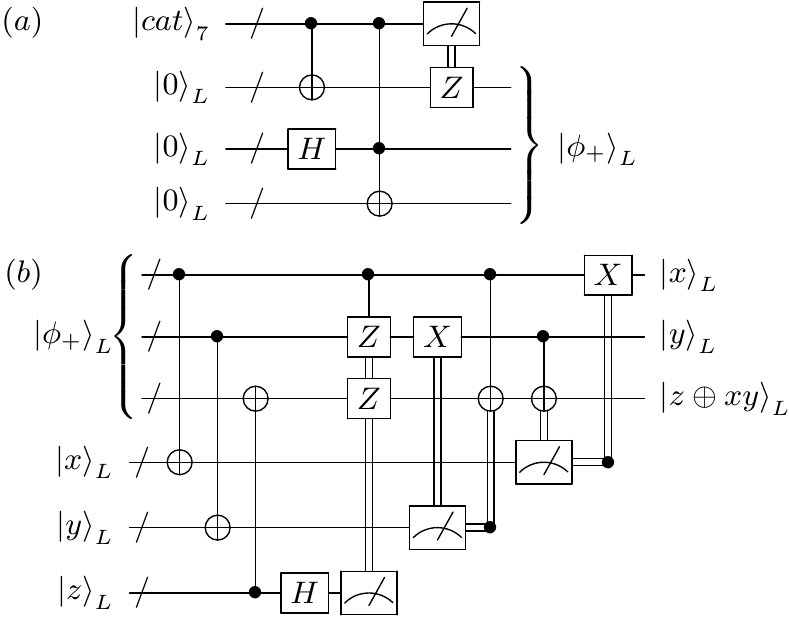}
\caption{Circuit diagram for realizing fault-tolerant Toffoli gate using Steane code. (a) The initial state $\ket{\phi_+}_L$ is prepared by measuring the $X_1$ and $CNOT_{12}$ of three qubit state $\ket{0}_1 (\ket{0}_2+\ket{1}_2) \ket{0}_3 / \sqrt{2}$. Note that the Toffoli gate shown here is a bitwise Toffoli between the 7-qubit cat state and the two logical qubit states. (b) Using the state prepared in (a), Toffoli gate can be realized using only measurement, Clifford group gates and classical communication, all of which can be implemented fault-tolerantly in the Steane code.}
\label{FTToffoli}
\end{figure}

\item In order to construct effective arithmetic circuits, we need Toffoli gate ({\em a.k.a.} $CCNOT_L$) which is not in the Clifford group. Since a transversal implementation of this gate is not possible in Steane code, fault-tolerant implementation requires preparation of a special three (logical) qubit state
\begin{equation}
\ket{\phi_+}_L = \frac{1}{2}(\ket{000}_L + \ket{010}_L + \ket{100}_L + \ket{111}_L),
\label{ToffoliState}
\end{equation}
and ``teleport'' the gate into this state \cite{ZhouPRA2000}. This state can be prepared by measuring its stabilizer operator using a 7-qubit cat state on three logical qubits $\ket{0}_L$, as shown in Fig. \ref{FTToffoli}a. Successful preparation of this state requires a bitwise Toffoli gate (at the physical level), which we assume can only be performed locally among qubits that are close to one another. Once this state is prepared, the three qubits $\ket{x}_L$, $\ket{y}_L$ and $\ket{z}_L$ participating in the Toffoli gate can be teleported to execute the gate, as shown in Fig. \ref{FTToffoli}b. Therefore, a successful Toffoli gate operation requires 3 logical qubits (which in turn require extra ancilla qubits to initialize) and 7 physical qubits as ancillary qubits, in addition to the three logical qubits on which the gate operates on.

\item When a $CNOT$ gate is necessary between two qubits that are separated by large distances, we take the approach where the two qubits of a maximally-entangled state is each distributed to the vicinity of the two qubits, and then the gate is teleported using the protocol proposed in Ref. \cite{GottesmanNature1999}. Efficient distribution of the entangled states makes this approach much more effective than where the qubits themselves are transported directly.

\end{enumerate}

\section{Error Probability for 3D Cluster States with Fast Entangling Gates}
\label{ErrorRateCalc}
Here we calculate the total error probability of the stabilizer measurement process for the model considered in Section~\ref{fastgates}, assuming independent strengths for the local errors and 2-local gate errors. We have local errors with strength $T/\tau_D$, and 2-local gate errors with strength $\epsilon$. The expectation value of the stabilizer operator $K_{\partial q}$ in Eq.~(\ref{Kcrit}) is
\begin{equation}
  \label{Kcrit2}
  \langle K_{\partial q}\rangle = \prod_{E \in \text{error sources}}1-2p_{E}.
\end{equation}
Therein, $p_{E}$ is the total probability of those Pauli errors in the error source $E$ which, after (forward) propagation to the endpoint of the cluster state creation procedure, anti-commute with the stabilizer operator $K_{\partial q}$. The r.h.s. of Eq.~(\ref{Kcrit2}) is simply a product due to the statistical independence of the individual error sources. Since the cluster state creation procedure is of bounded temporal depth and built of local and nearest-neighbor gates only, errors can only propagate a finite distance. Therefore, only a finite number of error sources contribute in Eq.~(\ref{Kcrit2}). 

To simplify the bookkeeping, we make the following observations. (a) A Bell state preparation, 2 CNOT gates (one on either side), and two local measurements on the qubits of the former Bell pair (one in the $Z$- and one in the $X$ basis) amount to a CNOT gate between remaining participating qubits. Therein, the qubit on the edge of the underlying lattice is the target, the qubit on the face is the control. We call this a teleported CNOT link. (b) Errors can only propagate once from face qubit to an edge qubit or vice versa, but never farther than that. To see this, consider e.g. a face qubit. There, an $X$- or $Y$-error can get propagated (face = control of CNOTs). In either case it causes an $X$-error on a neighboring edge qubit. But $X$-errors are not propagated from edge-qubits (edge = target of all CNOTs). (c) The stabilizer $K_{\partial q}$ has only support on face qubits, and is not affected by $X$-errors. 

Based on these observations, we subdivide the error sources affecting $\langle K_{\partial q}\rangle$ into three categories, namely Type 1: First Bell pair created on each face (according to the 5-step schedule); Type 2: The CNOT links, consuming the remaining Bell pairs; and Type 3: The final measurements of the cluster qubits (1 per ELU).    

Type-2 contributions: For every CNOT link we only need to count $Z$-errors (and $Y\cong Z$) on both the control (= face) and target (= edge), because on the face qubit the $Z$-errors are the ones that matter [with (c)], and on the edge qubit, such errors may still propagate to a neighboring face qubit [with (b)] and matter there. With these simplifications, the effective error of each CNOT link between two neighboring ELUs is described by the probabilities $p_{ZI}$ for a $Z$-error on the face qubit, $p_{IZ}$ for a $Z$-error on the edge qubit, and $p_{ZZ}$ for the combined error; and
\begin{equation}
\label{PzT2}
p_{ZI}=2\epsilon + \frac{10}{3}\frac{T}{\tau_D},\; p_{IZ}= p_{ZZ}=\frac{4}{15}\epsilon+\frac{2}{3}\frac{T}{\tau_D}.
\end{equation}
Herein, we have only kept contributions up to linear order in $\epsilon$, $T/\tau_D$. The contributions to the error come from (1) the Bell pair, (2) a first round of memory error on all qubits, (3) the CNOT gates, (4) a second round of memory error on all qubits, and (5) the two local measurements per link.

Now we need to discuss the effect of each of the above gates on $\langle K_{\partial q}\rangle$, taking into account propagation effects. For example, consider the link established between the face qubit of a front face $f$ with its left neighboring edge qubit. (The Bell pair for this link is created in Step 1, the required CNOTs are performed in Step 2, and the local measurements in Step 3.) The $Z$-error on $f$ does not propagate further. The $Z$-error on $e$ is propagated in later steps to a neighboring face (see Fig.~\ref{CluCr}b). Thus, the errors $Z_f$ and $Z_e$ of this gate affect $\langle K_{\partial q}\rangle$, and $Z_eZ_f$ doesn't. With Eq.~(\ref{Kcrit2}), the gate in question reduces  $\langle K_{\partial q}\rangle$ by a factor of $1 - 68/15\,\epsilon - 8 T/\tau_D$.

The following links contribute: three for every face in $\partial q$ from within the cell, and three more per face of $\partial q$ from the neighboring cells (links ending in an edge belonging to the cell $q$ can affect $\langle K_{\partial q}\rangle$ by propagation). (i) Contributions from within the cell. If a $Z_e$-error of the link propagates to an even (odd) number of neighboring faces in $q$, the total error probability affecting $\langle K_{\partial q}\rangle$ is $p_{ZZ}+p_{ZI}$ ($p_{IZ}+p_{ZI}$). But since $p_{1Z}=p_{ZZ}$, all 18 contributions from within the cell $q$ are the same, irrespective of propagation. (ii) Contributions from neighboring cells. Each of the 18 links in question contributes an effective error probability $p_{IZ}+p_{ZZ}$ if an error on the edge qubit of the link propagates to an odd number of face qubits in $\partial q$. By inspection of Fig.~\ref{CluCr}b, this happens for 6 links. With Eq.~(\ref{PzT2}), all the Type-2 errors reduce $\langle K_{\partial q}\rangle$ by a factor of
\begin{equation}
\label{Ty2c}
1-160\frac{T}{\tau_D}-88\epsilon.
\end{equation}

Type-1 contributions: Each of the initial Bell pair creations carries a two-qubit gate error of strength $\epsilon$, and memory error of strength $T/\tau_D$ on either qubit. Similar to the above case, we can group the 15 possible Pauli errors into the equivalence classes $I$, $Z_f$ ($Z_eZ_f\equiv I$ and $Z_e \equiv Z_f$ for Bell states). The single remaining error probability, for $Z_f$, is
\begin{equation}
\label{PzT1}
p_{ZI} = \frac{8}{15}\epsilon +\frac{4}{3}\frac{T}{\tau_D}. 
\end{equation}
For each face of $\partial q$, there is one Bell pair within the face that reduces $\langle K_{\partial q}\rangle$ by a factor of $1-2p_{ZI}$. Bell pairs from neighboring cells do not contribute an error here. Thus, all the Type-1 errors reduce $\langle K_{\partial q}\rangle$ by a factor of
\begin{equation}
\label{Ty1c}
1-8\frac{T}{\tau_D}-\frac{16}{5}\epsilon.
\end{equation}
Again, only the contributions to linear order in $\epsilon$, $T/\tau_D$ were kept.

Type-3 contributions: The only remaining error source is in the measurement of the one qubit per ELU which is part of the 3D cluster state. The strength of the effective error on each face qubit is $p_Z=2/3\,\epsilon$. Each of the six faces in $\partial q$ is affected by this error. Thus, all the Type-3 errors reduce $\langle K_{\partial q}\rangle$ by a factor of
\begin{equation}
\label{Ty3c}
1-8\epsilon.
\end{equation}  

Combining the contributions Eq.~(\ref{Ty2c}), (\ref{Ty1c}), (\ref{Ty3c}) of error Types 1 - 3 yields 

\begin{equation}
  \label{Keval}
  \langle K_{\partial q}\rangle = 1 - \frac{512}{5} \epsilon - 176\frac{T}{\tau_D}
\end{equation}
for the expectation value $\langle K_{\partial q}\rangle$.  


\end{document}